\documentclass[12pt,draftcls,onecolumn]{IEEEtran}

\usepackage[latin1]{inputenc}
\usepackage[T1]{fontenc}
\usepackage[english]{babel}
\usepackage{amssymb,amsmath,amsfonts}
\usepackage{times}
\usepackage{graphicx}
\usepackage{color}
\usepackage{verbatim}
\usepackage{cite}

\usepackage{subfig}		
\usepackage{pgf}		
\usepackage{tikz}
\usetikzlibrary{decorations}
\usetikzlibrary{patterns}
\usetikzlibrary{positioning}
\usetikzlibrary{backgrounds}

\usepackage{stmaryrd}

\usepackage{rotating}

\newtheorem{remark}{Remark}
\newtheorem{prop}{Proposition}
\newtheorem{theo}{Theorem}
\newtheorem{lemma}{Lemma}
\newtheorem{coro}{Corollary}
\newtheorem{assum}{Assumption}

\newcommand{\tr}[1]{\textrm{Tr}\left(#1\right)}

\newcommand{\ess}{\mathrm{ess}}
\newcommand{\floor}[1]{\left\lfloor #1 \right\rfloor}
\newcommand{\ceil}[1]{\left\lceil #1 \right\rceil}
\newcommand{\diag}[1]{\textrm{Diag}\left(#1\right)}
\renewcommand{\i}{\imath}

\providecommand{\norm}[1]{\left\lVert#1\right\rVert}
\newcommand{\abs}[1]{\left\lvert#1\right\lvert}

\newcommand{\sS}{{\mathsf S}}
\newcommand{\sX}{{\mathsf X}}
\newcommand{\sY}{{\mathsf Y}}

\newcommand{\cT}{{\mathcal T}}
\newcommand{\cU}{{\mathcal U}}
\newcommand{\cV}{{\mathcal V}}
\newcommand{\cW}{{\mathcal W}}

\newcommand{\bP}{{\mathbb P}}
\newcommand{\bR}{{\mathbb R}}
\newcommand{\bE}{{\mathbb E}}
\newcommand{\bZ}{{\mathbb Z}}

\newcommand{\HO}{{\mathsf H0}}
\newcommand{\HU}{{\mathsf H1}}

\newcommand{\cL}{{\mathcal L}}
\newcommand{\cH}{{\mathcal H}}

\newcommand{\cN}{{\mathcal{N}}}

\newcommand{\dron}[2]{\frac{\partial #1}{\partial #2}}
\newcommand{\dddron}[3]{\frac{\partial^2 #1}{\partial #2 \:\partial #3}}
\newcommand{\dtron}[4]{\frac{\partial^3 #1}{\partial #2 \:\partial #3\:\partial #4}}

\newcommand{\nabl}[1]{\nabla\!\!\!\:_{#1}}
\newcommand{\hess}[1]{\nabla\!\!\!\:_{#1}^{\;2}}

\newcommand{\cC}{{\mathcal C}}

\newcommand{\Nj}{_{N,j}}
\newcommand{\Nk}{_{N,k}}
\newcommand{\Nl}{_{N,\ell}}
\newcommand{\Njk}{_{N,j_k}}
\newcommand{\Nji}{_{N,j_i}}
\newcommand{\Njl}{_{N,j_\ell}}
\newcommand{\Njr}{_{N,j_r}}

\newcommand{\uN}{_{1,N}}
\newcommand{\oN}{_{0,N}}
\newcommand{\iN}{_{i,N}}

\newcommand{\T}{^\text{\scshape{t}}}

\newcommand{\ind}{{\mathbf 1}}

\newcommand{\intm}{\int\!\!\dots\!\!\int}

\title{High-Rate Vector Quantization for the Neyman-Pearson Detection of Correlated~Processes}

\author{
 	Joffrey Villard,~\IEEEmembership{Student~Member,~IEEE,} and
 	Pascal Bianchi,~\IEEEmembership{Member,~IEEE}
 	\thanks{J. Villard is with the Department of Telecommunications, SUPELEC, 91192 Gif-sur-Yvette, France (e-mail: joffrey.villard@supelec.fr).}
 	\thanks{P. Bianchi is with Telecom ParisTech, 75634 Paris Cedex 13, France (e-mail: bianchi@telecom-paristech.fr).}
 	\thanks{The work of J. Villard is supported by DGA (French Armement Procurement Agency).}
 }
	
\date{May 2011}

\begin{document}

\maketitle

\begin{abstract}
  This paper investigates the effect of quantization on the
  performance of the Neyman-Pearson test. 
  It is assumed that a sensing
  unit observes samples of a correlated stationary ergodic
  multivariate process. Each sample is passed through an $N$-point
  quantizer and transmitted to a decision device which performs a
  binary hypothesis test.  For any false alarm level, it is shown that
  the miss probability of the Neyman-Pearson test converges to zero
  exponentially as the number of samples tends to infinity, assuming that
  the observed process satisfies certain mixing conditions.
  The main contribution of this paper is to provide a compact closed-form
  expression of the error exponent in the high-rate regime
  \emph{i.e.}, when the number~$N$ of quantization levels tends to
  infinity, generalizing previous results of Gupta and Hero to the case 
  of non-independent observations. If $d$ represents the dimension of one sample, 
  it is proved that the error exponent converges at rate $N^{2/d}$
  to the one obtained in the absence of quantization.
  As an application, relevant high-rate quantization strategies which lead to a large error exponent are determined. 
  Numerical results indicate that the proposed quantization rule can yield better performance than existing ones in terms of detection error.
\end{abstract}


\section{Introduction}

Consider a sensing unit which
transmits a sequence of measurements to a decision device~(DD)
whose mission is to detect a given signal.
For example, a CCTV camera in a surveillance system transmits its data
to a remote controller interested in the detection of a particular
object in its field of view.  This situation also arises in the
context of wireless sensor networks (WSN) where a fusion center
collects the individual measurements of a large number of identical sensors
and processes these measurements in order to detect abnormal events
\cite{akyildiz2002wireless,chen2006channel}.  In such
applications, due to bandwidth, delay or storage limitations,
transmitted data rates are often limited.  Therefore, measurements must be 
quantized prior to transmission. As a matter of fact, this quantization step may severely 
degrade the overall detection performance of the system. 

In this paper, we consider that
a binary hypothesis test is performed at the DD. The available data set 
corresponds to a quantized version of a stationary
ergodic discrete-time multivariate process. 
Our aim is to quantify the detection performance of a given quantizer and
characterize quantization strategies which guarantee attractive performance at the DD.

In the past decades, numerous papers were dedicated to the search for relevant quantization
strategies and their practical design~\cite{gray1998quantization}.
The most popular criterion used to select quantizers is the mean square error (MSE) 
between the quantized signal and the initial source~\cite{gersho1992vector}.
An analytical characterization of quantizers minimizing the MSE is
difficult in the general case.  Bennett~\cite{bennett1948spectra}
pioneered the study of \emph{high-rate} (or \emph{high-resolution})
quantization for the reconstruction of scalar signals.  The idea of
Bennett was to study the MSE in the asymptotic regime where the number
of quantization levels tends to infinity. A closed form expression of
the (properly normalized) MSE can be determined in that case, and the
families of quantizers minimizing the asymptotic MSE can be directly
characterized.  Extension of the work of Bennett to vector-valued
observations was later achieved in~\cite{na1995bennett}.  However, the
MSE criterion is especially relevant when the aim is to reconstruct
the source.  On the other hand, it can be inappropriate as far as other
applications are concerned. For this reason, various distortion
measures have been proposed in the literature in a
\emph{task-oriented} setting for estimation, classification and
detection~\cite{han1998statistical,
  misra2008distributed,xiao2006distributed,
  perlmutter1996bayes,kassam1977optimum,poor1977applications,poor1988fine,picinbono1988optimum,
  tsitsiklis1993extremal,tenney1981detection,tsitsiklis1988decentralized,gupta2003high}.
In particular, considerable attention has been paid to optimal
quantization for hypothesis testing. 
Poor and Thomas~\cite{poor1977applications} used Ali-Silvey distances
between densities.  Later, Poor~\cite{poor1988fine} proposed the
generalized $f$-divergence and studied this distortion measure in the
high-rate regime.  Picinbono and Duvaut~\cite{picinbono1988optimum}
considered a deflection criterion and proved that the corresponding
optimal procedure corresponds to the scalar quantization of the
likelihood ratio.  Tsitsiklis~\cite{tsitsiklis1993extremal} studied
the properties of such quantizers with respect to several distortion
measures.  More recently, following the initial works of Tenney and
Sandell~\cite{tenney1981detection} and
Tsitsiklis~\cite{tsitsiklis1988decentralized}, Gupta and
Hero~\cite{gupta2003high} investigated the selection of high-rate
quantizers for binary hypothesis tests. In their setting, the decision
device gathers a sequence of $n$ independent and identically
distributed (i.i.d.) variables, each of these variables being passed
through a fixed quantizer.  The probability density function (pdf) of
the samples is assumed to be known both under the null hypothesis and
the alternative. In this case, it is well known that a uniformly most
powerful test is obtained by the Neyman-Pearson (NP) procedure which
consists in rejecting the null hypothesis when the 
log-likelihood ratio~(LLR) exceeds a certain
threshold~\cite{lehmann2005testing}. The threshold is usually chosen
in such a way that the probability of false alarm of the test (that
is, the probability to decide the alternative under the null
hypothesis) is fixed to a specified \emph{level}, say $\alpha$.  The
performance of the NP test of level $\alpha$ can be evaluated in terms of
the miss probability (that is, the probability to decide the null
hypothesis under the alternative).  In our case, the miss probability clearly depends on the
quantizer used by the sensing unit. Thus, a natural approach would be
to select the quantizer which minimizes the miss probability.
Unfortunately, the miss probability does not admit any tractable
expression as a function of the quantizer. To circumvent this issue,
it is convenient to study the miss probability in the case where the
number $n$ of available snapshots tends to infinity.  
In case of i.i.d. observations, 
the celebrated Stein's lemma~\cite{cover2006elements} states that the miss probability tends to zero
exponentially in $n$. Based on this result, it is relevant to select the quantizers
which yield a large value of the error exponent. Unfortunately, the
maximization of the error exponent as a function of the quantizer is
impractical. Following the idea of~\cite{bennett1948spectra,na1995bennett}, Gupta
and Hero restrict their attention to high-rate quantizers and manage
to obtain a compact expression of the error exponent loss induced by
quantization.

Most of these works address the case where observations are independent random variables. 
However, the detection of a correlated process is a crucial issue 
in many applications~\cite{chamberland2006dense,willett2000good,sung2006neyman,hachem2009error}.
In this case, fewer results are available in the literature.
Chamberland and Veeravalli~\cite{chamberland2006dense}
analyze the impact of the density of sensors in a WSN on the detection performance,
when observations are correlated.
Willett~\emph{et al.}~\cite{willett2000good} study the one-bit quantization 
of a pair of dependent Gaussian random variables.
In case of the detection of a Gauss-Markov signal in noise,
Sung~\emph{et al.}~\cite{sung2006neyman} prove that for a fixed false alarm level,
the miss probability of the NP test converges exponentially to zero,
and provide a closed form expression of the error exponent. 
Hachem~\emph{et al.}~\cite{hachem2009error} later extended the results of \cite{sung2006neyman}
to irregularly sampled Gaussian diffusion processes.
However, \cite{sung2006neyman,hachem2009error} assume that the DD has a perfect access to the observations
of the sensing unit, and do not address quantization issues.

In this paper, we study the performance of the Neyman-Pearson test based
on a quantized version of a stationary ergodic multivariate process.
We generalize the work of Gupta and Hero~\cite{gupta2003high} to the case
where the observed process is non-i.i.d. (either under the null hypothesis, the alternative, or both).
In this situation, Stein's lemma does not directly apply. The  error
exponent does no longer admit a closed-form expression and the determination
of relevant quantizers is therefore a more difficult task.
Provided that the process of interest satisfies certain forgetting properties
(present observations should become nearly independent of past observations after a sufficient amount of time),
we prove that the miss probability of the NP test of level $\alpha$ tends exponentially to zero
as the number of observations tends to infinity.
Our main contribution is to provide a compact closed form expression
of the error exponent in case of high-rate quantizers.  If $N$ denotes
the number of quantization levels (or equivalently if each measurement
is quantized on $\log_2(N)$ bits), we prove that the error exponent
achieved when using quantized observations converges as $N$ tends to
infinity to the ideal error exponent that one would obtain if
perfect/unquantized measurements were available at the DD. More
precisely, we prove that the error exponent loss tends to zero at
speed $N^{-2/d}$ where $d$ represents the dimension of each individual
measurement.  
The asymptotic error exponent depends on the process
distributions under both hypotheses. It also depends on the
quantization strategy through the so-called \emph{model point density}
and \emph{model covariation profile}. The model point density can be
interpreted as the asymptotic density of cells in the neighborhood
of each point of the observation space. The model covariation profile
captures the shape of the cells.  As a consequence, the selection
of relevant high-rate quantizers reduces to the determination of the
point densities and covariation profiles minimizing the asymptotic
error exponent loss. In case of scalar quantization ($d=1$), our
compact expression immediately yields a simple characterization of
optimal high-rate quantizers.  In case of vector quantization ($d\geq
2$), an exact characterization of optimal quantizers is more
difficult. Following the approach of~\cite{gupta2003high} once again,
we nevertheless determine relevant families of quantizers with
attractive error exponent.  Note that our theoretical results hold
under the assumption that the observed process ``forgets'' past
observations fast enough.  As a special case, we prove that our
assumptions hold for a general class of hidden Markov
models verifying a certain contraction property.  Numerical
illustrations are provided in the case where the measurements correspond to a modulated signal in the In-phase/Quadrature plane.

The paper is organized as follows.
In Section~\ref{sec:framework}, we describe the observation model.
We also review some known results on Neyman-Pearson tests
and we derive the associated error exponent in the ideal case
where the DD has perfect access to the measurements.
The vector quantization framework is introduced in Section~\ref{sec:quantized}.
In Section~\ref{sec:high-rate}, the impact of quantization on the error 
exponent is evaluated in the high-rate regime.
We determine relevant quantization strategies
allowing to reduce this degradation.
Section~\ref{sec:proof} is devoted to the proof of the main result.
In Section~\ref{sec:illus}, we illustrate our findings
in the special case of hidden Markov processes 
and give sufficient conditions on the transition and observation kernels 
ensuring that our results apply.
Section~\ref{sec:numerical} is dedicated to numerical illustrations.

\subsection*{Notation}

For any sequence~$(y_i)_{i\in\bZ}$,  for any integers~$k\leq\ell$, notation~$y_{k:\ell}$ stands for the collection $(y_k,y_{k+1},\dots, y_\ell)$
and notation $y_\bZ$ is used to designate the whole sequence.
If $y$ is a vector with dimension~$d$, we denote by $y^{(i)}$ its $i$-th component and $\norm{y}$ its Euclidean norm. 
We denote by $\norm{A}$ the spectral norm of any square matrix~$A$.
Notation~$.\T$ stands for the transpose operator.

A real-valued function $f: y_{k:\ell} \mapsto f(y_{k:\ell})$ on $\sS\subset\bR^d× \cdots× \bR^d$ is said to be of class $C_3$ on $\sS$ if it is three times continuously differentiable on $\sS$. We denote by $\nabl{y_m}f(y_{k:\ell})$ its gradient w.r.t.~$y_m$ at point~$y_{k:\ell}$. When no variable is specified, $\nabla g(y)$ simply denotes the ($d$-dimensional) gradient of the real-valued single-variable function $y \mapsto g(y)$ defined on $\sY\subset\bR^d$.
We define the Hessian matrix of~$f$ by $\left[\hess{y_m,y_n}f\right]_{i,j} = \dddron{f}{y_m^{(i)}}{y_n^{(j)}}$ for all $i,j\in\{1,\dots,d\}$. Moreover, notation $\hess{y_m}$ stands for $\hess{y_m,y_m}$. 

Notation $B(\sX)$ stands for the Borel $\sigma$-field on $\sX$.
Notation $\sigma(Y_{1:n})$ stands for the sub-$\sigma$-field of $B(\sY^\bZ)$, associated with the random vector $Y_{1:n}$. 
Notation $\xrightarrow[n\to\infty]{P}$ stands for the convergence in probability as $n\to\infty$.
Notation $\xrightarrow[n\to\infty]{L^r(\bP_0)}$ stands for the convergence in the $L^r$-norm w.r.t. probability $\bP_0$.
 
Notation $\circ$ stands for the composition operator \emph{i.e.}, for any arbitrary functions $f$ and $g$, $f\circ g(x) = f(g(x))$.
Notation $o_N(\cdot)$ is a little-o notation as $N$ tends to infinity.

\section{Neyman-Pearson Detection with Perfect Observations}
\label{sec:framework}

\subsection{Observation Model}
\label{sec:model}

Consider two probability measures $\bP_0$ and $\bP_1$ on a relevant
probability space.  
Denote by $(Y_k)_{k\in \bZ}$ a stationary ergodic process
for both $\bP_0$ and $\bP_1$, taking its values in a bounded 
convex subset $\sY$ of~$\bR^d$. 
We associate an hypothesis ($\HO$ and $\HU$ respectively) to each of
the two probability measures $\bP_0$ and $\bP_1$ and investigate the
problem of the detection of $\HU$ \emph{vs.} $\HO$ based on a set of
$n$ observations $Y_{1:n} = (Y_1,\dots, Y_n)$.

For each $i\in\{0,1\}$, we assume that $\bP_i$ is the probability
distribution of the coordinate process $(Y_k)_{k\in \bZ}$ on the
canonical space $(\sY^\bZ,B(\sY^\bZ))$. We denote by $P_{i,n}$ the
restriction of $\bP_i$ to $\sigma(Y_{1:n})$.  We denote by $\bE_0$ and
$\bE_1$ the expectations associated with $\bP_0$ and $\bP_1$
respectively.  We introduce the reference measure $\mu$ which coincides
with the $d$-dimensional Lebesgue measure restricted to~$\sY$.
\begin{assum}
\label{ass:dens}
The following properties hold true for each $i\in\{0,1\}$.
\begin{enumerate}
\item For each $n\geq 1$, $P_{i,n}$ admits a density
$p_{i}$ w.r.t.~$\mu^{\otimes\,n}$.
\item $p_{i}(y_{1:n}) > 0$ for each $y_{1:n}\in \sY^{n}$.
\item $\bE_0\left|\log p_i(Y_0)\right| < \infty$.
\end{enumerate}
\end{assum}
The density $p_i$ of $P_{i,n}$ depends of course on $n$, but we drop
the index $n$ to simplify the notation.  For each $i\in\{0,1\}$, we also
define $p_i(y_n| y_{1:n-1})= {p_i(y_{1:n})}/{p_i(y_{1:n-1})}$ with the
convention that $p_i(y_n| y_{1:n-1})=p_i(y_n)$ when $n=1$ (that is,
when $y_{1:n-1}$ is a void vector).  
Assumption~\ref{ass:dens}-2) implies that both distributions $P_{0,n}$
and $P_{1,n}$ are absolutely continuous w.r.t. each other.

\subsection{Likelihood Ratio Test}

We now investigate the detection of $\HU$ \emph{vs.} $\HO$
based on the perfect observation of $n$ measurements $Y_{1:n}$.
The log-likelihood ratio (LLR) writes:
\begin{equation}
\label{eq:llr}
L_n = \log\frac{p_1(Y_{1:n})}{p_0(Y_{1:n})}\ .
\end{equation}
The NP test rejects the null hypothesis when $L_n$ is larger than a threshold, say $\gamma$.
For each $\alpha\in (0,1)$, we define the miss probability of the NP test of level $\alpha$ by:
\[
  \beta_n(\alpha) = \inf  \bP_1\left[ L_n < \gamma\right]\ ,
\]
where the infimum is  w.r.t. all $\gamma$ such that the probability of false alarm
does not exceed $\alpha$ \emph{i.e.},
$$
\gamma\textrm{ s.t. }\bP_0\left[ L_n > \gamma\right] \leq \alpha\ .
$$
For each $n\geq 1$ and each $\alpha\in(0,1)$, due to the celebrated Neyman-Pearson's lemma, $\beta_n(\alpha)$ is the lowest achievable
miss probability among all binary tests of level $\alpha$ which are based on the observation
of $Y_{1:n}$.
Quantity $\beta_n(\alpha)$ is therefore a key metric in order to characterize the performance of the hypothesis test.
Unfortunately, it usually does  not admit any tractable closed-form expression.
In the sequel, we study the asymptotic behaviour of $\beta_n(\alpha)$ as the number of observations
$n$ tends to infinity. In this regime, it can be shown that, under certain assumptions,
\begin{equation}
\label{eq:errexp_rough}
\beta_n(\alpha)\simeq \exp(-n\,K)
\end{equation}
for some constant $K$ given below, which we shall refer to as the error exponent.

\subsection{Error Exponent with Perfect Observations}

The evaluation of the error exponent $K$ in Equation~(\ref{eq:errexp_rough})
fundamentally relies on the following lemma:
\begin{lemma}[\cite{chen1996general}]
\label{lem:chen}
Assume that a binary test is performed on a sequence $\check Y_{1:n} = (\check Y_1,\dots,\check Y_n)$ 
of $n$ observed random variables. 
Denote by $\check p_0$ and $\check p_1$ the density of $\check Y_{1:n}$ under $\HO$ and $\HU$ respectively
(w.r.t. any common reference measure).
Assume that under $\HO$,
$$
\frac 1n \log\frac{\check p_0(\check Y_{1:n})}{\check p_1(\check Y_{1:n})} \xrightarrow[n\to\infty]{P} \kappa 
$$
for some deterministic constant $\kappa$ such that $0<\kappa\leq \infty$. 
Then, for any $\alpha\in(0,1)$ the miss probability $\beta_n(\alpha)$ of the Neyman-Pearson test of level $\alpha$
is such that
\[
\lim_{n\to\infty} \frac 1n \log\beta_n(\alpha) = -\kappa\ .
\]
\end{lemma}

Lemma~\ref{lem:chen} implies that the error exponent, if it exists, coincides
with the limit in probability (under~$\bP_0$) of~$-(1/n) L_n$, 
where $L_n$ is the LLR defined by~(\ref{eq:llr}).
The existence of the error exponent is directly obtained from the following assumption,
which will be discussed later on.

\begin{assum}
\label{ass:logp}
For each $i\in\{0,1\}$, $(\log p_i(Y_0|Y_{-m:-1}))_{m\geq 0}$ is a convergent sequence in $L^1(\bP_0)$.
\end{assum}

We are now in position to study the limit of the LLR $L_n$
and prove the following result, 
which provides the general form of the error exponent.

\begin{theo}
\label{th:exp_err}
Under Assumptions~\ref{ass:dens} and~\ref{ass:logp},  
$$
\lim_{n\to\infty} \frac 1n \log\beta_n(\alpha) = -K\ ,
$$
where $K$ is the constant defined by
\begin{equation}
\label{eq:K}
K = \lim_{m\to\infty} \bE_0\left[\log \frac{p_0}{p_1}(Y_0|Y_{-m:-1})\right]\ .
\end{equation}
\end{theo}

\begin{IEEEproof}
Using the chain rule, we first write $L_n$ under the form:
\[
L_n =  -\sum_{k=1}^n \log \frac{p_0}{p_1}(Y_k | Y_{1:k-1})\ .
\]
Denote by $\Upsilon$ the limit in $L^1(\bP_0)$ of sequence $(\log \frac{p_0}{p_1}(Y_0|Y_{-m:-1}))_{m\geq 0}$.
The main point is the study of the difference $\log \frac{p_0}{p_1}(Y_k | Y_{1:k-1}) -\Upsilon\circ\theta^k$,
where $\theta$ is the shift operator\footnote{Recall that we are considering probability measures
defined on the canonical space $\sY^\bZ$. For any $\omega\in\sY^\bZ$, we may write
$\omega=(\dots,\omega_{-1},\omega_0,\omega_1,\dots)$. The $k$th-time shifted version of $\omega$ is then given by
$\theta^k \omega=(\dots,\omega_{k-1},\omega_k,\omega_{k+1},\dots)$. Notation $\Upsilon\circ\theta^k$ represents the measurable function
$\Upsilon\circ\theta^k(\omega)=\Upsilon(\theta^k \omega)=\Upsilon((\dots,\omega_{k-1},\omega_k,\omega_{k+1},\dots))$.
Recall that process $Y_\bZ$ is defined as the coordinate process \emph{i.e.}, $Y_n(\omega) =
\omega_n$ for each $n$. As a consequence, the measurable function
$\log \frac{p_0}{p_1}(Y_k | Y_{1:k-1}) -\Upsilon\circ\theta^k$ at point $\omega$ is equal to the measurable function
$\log \frac{p_0}{p_1}(Y_0 | Y_{-k+1:-1}) -\Upsilon$ at point $\theta^k\omega$.
}.
We can write:
\begin{eqnarray*}
\bE_0\left|\frac1n\,L_n + \frac1n\sum_{k=1}^n \Upsilon\circ\theta^k\right|
	&\stackrel{(a)}{\leq}& \frac1n\,\sum_{k=1}^n \bE_0\left|\log \frac{p_0}{p_1}(Y_k | Y_{1:k-1}) -\Upsilon\circ\theta^k\right| \\
	&\stackrel{(b)}{\leq}& \frac1n\,\sum_{k=1}^n \bE_0\left|\log \frac{p_0}{p_1}(Y_0 | Y_{-k+1:-1}) -\Upsilon \right|\ ,
\end{eqnarray*}
where step~$(a)$ comes from the triangular inequality
and step~$(b)$ is a consequence of the stationarity of process $(Y_k)_{k\in \bZ}$ under $\bP_0$.
The right-hand side of the above inequality can be interpreted as a Cesàro mean
and thus converges to zero by definition of $\Upsilon$.
We thus write:
$$
-\frac 1n L_n = \frac 1n \sum_{k=1}^n \Upsilon\circ\theta^k + \varepsilon_n \ ,
$$
where $\varepsilon_n$ represents a term which converges in probability (under~$\bP_0$) to zero as $n\to\infty$.
As $\bP_0$ is stationary ergodic, we conclude using the ergodic theorem that
$-(1/n)L_n$ converges in probability to $\bE_0(\Upsilon)$ under $\bP_0$. 
This result together with Lemma~\ref{lem:chen} proves Theorem~\ref{th:exp_err}.
\end{IEEEproof}

\begin{remark}
  Let us make some remarks on the above Assumptions~\ref{ass:dens}
  and~\ref{ass:logp}.  Assumption~\ref{ass:dens} is an extension of
  those made by Gupta and Hero~\cite[Section~III, pp.
  1956]{gupta2003high}.  Assumption~\ref{ass:logp} does not appear
  in~\cite{gupta2003high} since it is obviously verified by i.i.d.
  processes. In this case, Theorem~\ref{th:exp_err} is known as
  Stein's lemma.  Assumption~\ref{ass:logp} is trivially satisfied by
  short-dependent ($m$-dependent) processes such as moving average
  processes for instance~\cite{bradley2005basic}.  In this case, the
  present observation $Y_0$ is independent of past observations
  $Y_{-m-1},Y_{-m-2},\dots$ as soon as $m$ is large enough.  As
  explained in Section~\ref{sec:illus}, Assumption~\ref{ass:logp} is
  as well satisfied by a wide class of hidden Markov models.
\end{remark}

\begin{remark}
In order that $(\log p_0(Y_0|Y_{-m:-1}))_{m\geq 0}$ is a convergent sequence in $L^1(\bP_0)$,
it is sufficient to check that 
 $(\bE_0\log p_0(Y_0|Y_{-m:-1}))_{m\geq 0}$ is a bounded sequence.
This claim is a consequence of Moy~\cite{moy1961generalizations} (see Theorem~4 therein).
In practical situations, this remark provides us with a convenient way to check 
whether Assumption~\ref{ass:logp} is verified for $i=0$.
On the other hand, the validation of Assumption~\ref{ass:logp} for $i=1$ 
generally requires more efforts in practice: One should be able
to prove that $(\log p_1(Y_0|Y_{-m:-1}))_{m\geq 0}$ is a Cauchy sequence in $L^1(\bP_0)$.
\end{remark}
\begin{remark}
When $\bP_1$ is a finite-order Markovian measure, 
Assumption~\ref{ass:logp} can be simply reduced to the assumption that
sequence $(\bE_0\log \frac{p_0}{p_1}(Y_0|Y_{-m:-1}))_{m\geq 0}$ is bounded. 
Indeed, due to Moy~\cite{moy1961generalizations}, this hypothesis directly implies the convergence 
of sequence $(\log \frac{p_0}{p_1}(Y_0|Y_{-m:-1}))_{m\geq 0}$ in $L^1(\bP_0)$ 
and thus yields Theorem~\ref{th:exp_err}.
\end{remark}

\section{Quantization}
\label{sec:quantized}

\subsection{Definitions}
\label{sec:quantif_def}

Consider a fixed integer $N\geq 2$. An $N$-point quantizer is a
triplet~$(\cC_N,\Xi_N,\xi_N)$ where
$\cC_N=\{C_{N,1},\dots,C_{N,N}\}$ is a set of $N$ \emph{cells}
(Borel sets of $\sY$ with non-zero volume) which form a partition
of $\sY$, where $\Xi_N=\{\xi_{N,1},\dots,\xi_{N,N}\}$ is an arbitrary set of
distinct elements and where $\xi_N:\sY\to\Xi_N$ is a function s.t. $\xi_N(y) =
\xi\Nj$ whenever $y\in C\Nj$. For each $N,k$, we introduce
\[
Z\Nk = \xi_N(Y_k) \ ,
\]
the quantized measurement on $\log_2(N)$ bits.  
We assume that the quantizer $(\cC_N,\Xi_N,\xi_N)$ is known at
the decision device. The aim is to decide between hypotheses $\HO$ and
$\HU$ based on the observation of $Z_{N,1:n}$.  

Note that in our model, each individual measurement is quantized 
based on the same quantization rule as in the traditional
framework of vector-quantization~\cite{gray1998quantization}.
It is also relevant in the case of WSN when samples are collected using identical
sensors.

\subsection{Error Exponent}

Assume that the number of quantization levels $N$ is fixed. 
For a given number $n$ of quantized observations, we define the LLR 
based on quantized measurements by:
\[
L_{n,N} = \log \frac{p\uN\left(Z_{N,1:n}\right)}{p\oN\left(Z_{N,1:n}\right)}\ ,
\]
where for  each $i\in\{0,1\}$ and for any set of quantization points 
$\xi_{N,j_{1:n}} = (\xi_{N,j_1},\dots,\xi_{N,j_n})\in\Xi_N^n$,
\[
p\iN(\xi_{N,j_{1:n}}) = P_{i,n}(C_{N,j_1}× \dots× C_{N,j_n})
\]
is the probability that measurements $Y_1,\dots,Y_n$ respectively fall 
into the cells $C_{N,j_1}, \dots, C_{N,j_n}$ associated with the observed
points $\xi_{N,j_1},\dots,\xi_{N,j_n}$ (\emph{n.b.} function $p\iN$ depends on $n$, but we omit the index
$n$ to simplify notation). 
We define similarly:
$$
p\iN(\xi_{N,j_n}|\xi_{N,j_{1:n-1}}) = \frac{p\iN(\xi_{N,j_{1:n}})}{p\iN(\xi_{N,j_{1:n-1}})}\ .
$$

For each $\alpha\in(0,1)$, we denote by $\beta_{n,N}(\alpha)$ the miss probability of the NP test of level $\alpha$
when quantization is applied \emph{i.e.}, 
the infimum of $\bP_1\left[L_{n,N}<\gamma\right]$ w.r.t. all $\gamma$ s.t. $\bP_0\left[L_{n,N}>\gamma\right]\leq \alpha$.
The error exponent associated with $\beta_{n,N}(\alpha)$ is provided by the following result,
whose proof is similar to the one of Theorem~\ref{th:exp_err}.
\begin{coro}
\label{coro:KN}
Consider a fixed $N\geq 2$. If Assumption~\ref{ass:dens} holds and
if   $(\log p\iN(Z_{N,0} | Z_{N,-m:-1}))_{m\geq 0}$ 
is a convergent sequence in $L^1(\bP_0)$ for each $i\in\{0,1\}$ then,
$$
\lim_{n\to\infty} \frac 1n\log\beta_{n,N}(\alpha) = -K_N\ ,
$$ 
where $K_N$ is the constant defined by:
\begin{equation}
  \label{eq:KN}
  K_N = \lim_{m\to\infty} \bE_0\left[\log \frac{p_{0,N}}{p_{1,N}}(Z_{N,0}|Z_{N,-m:-1})\right]\ .
\end{equation}
\end{coro}

The above result provides the error exponent $K_N$ associated with the NP test on quantized observations. A natural question is: How does the choice of the quantizer affect the error exponent? Unfortunately, the expression of the error exponent does not immediately allow to evaluate the impact of the quantizer.
In the sequel, we thus follow the approach of~\cite{na1995bennett,gupta2003high} and focus on the case where the order $N$ of the quantizer tends to infinity. We refer to such quantizers as \emph{high-rate} quantizers. 
This approach leads to a convenient and informative asymptotic expression of $K_N$.
In particular, it will be shown that, under some assumptions on the process $(Y_k)_{k\in\bZ}$ and the quantizers sequence $(\cC_N,\Xi_N,\xi_N)_{N\geq 1}$, the above error exponent $K_N$ converges to $K$ as $N$ tends to infinity.

\section{Performance of High-Rate Vector Quantizers}
\label{sec:high-rate}

\subsection{Notation and Assumptions}

For each $N$, we remark that the error exponent $K_N$ does not depend
on the particular choice of the quantization alphabet $\Xi_N$.\footnote{The value
  of the log-likelihood ratio (and \emph{a fortiori} the value of the
  error exponent) remains unchanged by any one-to-one transformation
  of the quantized observations. Otherwise stated, the particular definition of the quantization
alphabet has no impact on the corresponding Neyman-Pearson test provided that
the latter quantization alphabet is composed by $N$ distinct elements.}
For the sake of simplicity, 
we assume with no loss of generality that\footnote{The $i$th component of $\xi\Nj$ is defined as $\xi\Nj^{(i)} \triangleq  \left(\int_{C\Nj} y^{(i)}\,dy\right)/\left(\int_{C\Nj} dy\right)$.}:
\[
\xi\Nj = \frac{\int_{C\Nj} y\,dy}{\int_{C\Nj} dy}\ ,
\]
\emph{i.e.} each $\xi\Nj$ coincides with the centroid of cell $C\Nj$.
We respectively define the volume and the diameter of cell $j$ by $V\Nj = \int_{C\Nj} dy$
and $d\Nj = \sup_{u,v\in C\Nj} \norm{u-v}$.
We introduce the \emph{specific point density} $\zeta_N$ and the 
\emph{specific covariation profile} $M_N$ as the piecewise constant functions on $\sY$ respectively defined
as follows, for any $y\in C\Nj$ $(j\in\{1,\dots,N\}$):
\begin{align*}
\zeta_N(y)&\,=\,\zeta\Nj\,=\,\frac 1{N V\Nj} \ ,\\
M_N(y)&\,=\,M\Nj\,=\,\frac 1{V\Nj^{1+2/d}}\int_{C\Nj} (y-\xi\Nj)(y-\xi\Nj)\T dy  \ .
\end{align*}
Now consider a family of quantizers $(\cC_N,\Xi_N,\xi_N)_{N\geq 1}$. 
We make the following assumption.
\begin{assum}
\label{ass:high-rate}
The following properties hold true.
\begin{enumerate}
\item As $N\to\infty$, $\zeta_N$ converges uniformly to a 
continuous function $\zeta$ such that $\inf_{y\in \sY} \zeta(y)>0$~.
\item As $N\to\infty$, $M_N$ converges uniformly to a continuous (matrix-valued) function $M$
such that $\sup_{y\in \sY} \norm{M(y)}<\infty$~.
\item There exists a constant $C_d$ such that, for all $N$,
$\sup_j d\Nj\leq\dfrac{C_d}{N^{1/d}}$ .
\end{enumerate}
\end{assum}
We will refer to $\zeta$ as the \emph{model point density} of the family
$(\cC_N,\Xi_N,\xi_N)_{N\geq 1}$. It represents the fraction of cells in the
neighborhood of a given point $y$. Function $M$ will be referred to as
the \emph{model covariation profile}. For each $y\in\sY$, $M(y)$ is a
non-negative $d× d$ matrix. In the literature, function $y\mapsto \tr{M(y)}$
is usually referred to as the \emph{inertial
  profile}~\cite{gray1998quantization,na1995bennett,gupta2003high}.
Function $M$ provides information about the shape of the cells.

Intuitively, high-rate quantizers should be constructed in such a way
that $\zeta(y)$ is large at those points $y$ for which a fine quantization
is essential to discriminate the two hypotheses.
Theorem~\ref{th:cv_ordre2} below provides a more rigorous formulation
of this intuition.

\begin{remark}
  Assumption~\ref{ass:high-rate} is essentially the same as the one
  traditionally made in the \emph{high-rate quantization}
  framework~\cite{gray1998quantization,na1995bennett,gupta2003high}.
The main difference lies in Assumption~\ref{ass:high-rate}-3):
  Usually, the volume of each cell vanishes at speed $1/N$ while the
  diameter tends to zero. Our assumption introduces a constraint on the
  speed of convergence of the sequence of diameters $\{d\Nj\}$, which
  ensures that cells shrink at the same speed ($1/N^{1/d}$) on each
  dimension. Assumption~\ref{ass:high-rate} is for instance valid for sequence
of quantizers constructed as \emph{companders}~\cite{bennett1948spectra,gray1998quantization}.
Such quantizers write as the composition of an invertible function (the so-called \emph{compressor})
and a uniform quantizer. Since~\cite{bennett1948spectra}, it is known that any scalar quantizer can be
written as a compander. Under mild conditions on the compressor,
 it can be shown that any sequence of companders with a given fixed compressor
satisfies Assumption~\ref{ass:high-rate} (in this case, the model point density $\zeta$ is fully determined by the 
first order derivative of the compressor). This point is discussed in Section~\ref{sec:scalar}.
\end{remark}

\subsection{Error Exponent in the High-Rate Regime}

Before stating the main result, we need further assumptions.
For each $m\geq 0$ and each $i\in\{0,1\}$, define:
\begin{eqnarray}
	\eta_i(m) 		&=& \sup_{m'\geq m} \bE_0\left|\log p_i(Y_0|Y_{-m:-1})-\log p_i(Y_0|Y_{-m':-1})\right| \ ,
	\label{eq:eta}\\
	\eta_{i,N}(m) 	&=& \sup_{m'\geq m} \bE_0\left|\log p_{i,N}(Z_{N,0}|Z_{N,-m:-1})-\log p_{i,N}(Z_{N,0}|Z_{N,-m':-1})\right|\ 
	\nonumber
\end{eqnarray}
Note that we already assumed in Theorem~\ref{th:exp_err} and Corollary~\ref{coro:KN}
that sequences $\log p_i(Y_0|Y_{-m:-1})$
and $\log p_{i,N}(Z_{N,0}|Z_{N,-m:-1})$ converge in $L^1(\bP_0)$ as $m\to\infty$,
meaning that $\eta_i(m)$ and $\eta_{i,N}(m)$ tend to zero.
Now coefficients $\eta_i(m)$ and $\eta_{i,N}(m)$ characterize the speed at which 
$\log p_i(Y_0|Y_{-m:-1})$ and $\log p_{i,N}(Z_{N,0}|Z_{N,-m:-1})$ converge to their limits.
They are therefore related to the mixing property of processes $Y_\bZ$ and $Z_{N,\bZ}$
(this point is discussed below in Remark~\ref{rem:mixing}).
In the sequel, we will need to ensure that these limits are reached fast enough
(see Assumption~\ref{ass:loss}-3) below).

\begin{assum}
\label{ass:loss}
The following properties hold true.
\begin{enumerate}
\item For any $n\geq 1$, $y_{1:n} \mapsto p_i(y_{1:n})$ is of class~$C_3$ on $\sY^{n}$.
\item $\sup_{\{n\geq 1, y_{1:n}\in \sY^{n},\;1\leq k,\ell,r \leq n,\;1\leq h,\bar\i,\bar\jmath \leq d\}}
		\Big| \dtron{\log p_i}{y_k^{(h)}}{y_\ell^{(\bar\i)}}{y_r^{(\bar\jmath)}}(y_{1:n}) \Big| < \infty$ .
\item There exist two constants $C_e$, $\epsilon>0$ such that for each $i\in\{0,1\}$, $N\geq 2$ and  $m\geq 0$,
\begin{equation}
\label{eq:6+eps}
\max\left( \eta_{i}(m) , \eta_{i,N}(m) \right)\leq \frac{C_e}{(1+m)^{6+\epsilon}}\ .
\end{equation}
\item For each $i\in\{0,1\}$,
each integers $m,m',k$ such that~$-m'\leq-m\leq 0\leq k$:
  \begin{align}
    &\bE_0\norm{\nabl{y_0}\log{p_i}(Y_{0:k} | Y_{-m:-1})-\nabl{y_0}\log{p_i}(Y_{0:k} | Y_{-m':-1})} \leq \varphi_m \ ,
	\label{eq:phi}\\
    &\bE_0\norm{\nabl{y_0}\log{p_i}(Y_k|Y_{-m:k-1})} \leq \psi_k 
	\label{eq:psi}\ ,
  \end{align}
where $\sum_k\varphi_k$ and $\sum_k\psi_k$ are convergent series.
\end{enumerate}
\end{assum}

Assumption~\ref{ass:loss} will be discussed in details at the end of the present subsection.
Particular examples of processes satisfying the above assumption are provided in Section~\ref{sec:illus}
and in the numerical results of Section~\ref{sec:numerical}.
We are now in position to state our main result.
Recall that $p_0(y)$ is the pdf of $Y_0$ under $\bP_0$.
Recall that $K$ and $K_N$ are the error exponents associated with the NP test in the absence and in the presence of quantization respectively, given by~(\ref{eq:K}) and~(\ref{eq:KN}). 
Note that Assumption~\ref{ass:loss}-3) implies that both sequences $\eta_i(m)$ and $\eta_{i,N}(m)$ tend to zero. This guarantees that under Assumption~\ref{ass:dens} the conclusions of Theorem~\ref{th:exp_err} and Corollary~\ref{coro:KN} hold true \emph{i.e.}, error exponents $K$ and $K_N$ do exist.

\begin{theo}
\label{th:cv_ordre2}
Under Assumptions~\ref{ass:dens},~\ref{ass:high-rate},~\ref{ass:loss}, 
the following statement holds true:

As $N$ tends to infinity, $N^{2/d}(K-K_N)$ converges to a constant $D_e$ given by
\begin{equation}
\label{eq:loss}
D_e =\frac 1 2\, \int \frac{p_0(y)F(y)}{\zeta(y)^{2/d}}\,dy \ ,
\end{equation}
where function $F$ is given by
\begin{equation}
\label{eq:F}
F(y) = \bE_0 \left[\ell(Y_\bZ)\T M(Y_0)\, \ell(Y_\bZ)\,\Big|\,Y_0=y \right] ,
\end{equation}
and random variable $\ell(Y_\bZ)$ is the limit in $L^2(\bP_0)$ of sequence $\left( \nabl{y_0}\log\frac{p_0}{p_1}(Y_{-k:k}) \right)_{k\geq0}$.
\end{theo}

The proof of Theorem~\ref{th:cv_ordre2} is given in Section~\ref{sec:proof}.

Theorem~\ref{th:cv_ordre2} states that when the order of the quantizer
tends to infinity, the error exponent $K_N$ associated with the NP
test converges at speed $N^{-2/d}$ to the error exponent $K$
that one would have obtained in the absence of quantization.  Loosely
speaking, if $\beta_{n,N}(\alpha)$ represents the miss probability of the NP
test of level $\alpha$, the approximation
\begin{equation}
\label{eq:approx_betanN}
\beta_{n,N}(\alpha) \simeq e^{-n\left(K-\frac{D_e}{N^{2/d}}\right)}
\end{equation}
holds when both the number $n$ of sensors and the order $N$ of
quantization are large. Quantity $D_e$ represents the 
(normalized) loss in error exponent between the quantized and the unquantized cases,
in the high-rate quantization regime.

Note that Equation~\eqref{eq:loss} resembles to Bennett's formula~\cite[Equation (1.6)]{bennett1948spectra} and its vector extension for $r$th-power distortion~\cite[Equation (7)]{na1995bennett}. 

\begin{remark}
As a first consequence of Theorem~\ref{th:cv_ordre2}, under some assumptions on the process, \emph{classical} quantizers as those produced in an MSE perspective will lead to error exponent $K_N$ which converges to $K$ as $N$ tends to infinity, at speed $N^{-2/d}$ (see Equation~\eqref{eq:approx_betanN} above).
\end{remark}

\begin{remark}
  The particular situation where measurements $(Y_k)_{k\geq 0}$ are
  i.i.d. under both hypotheses was studied by Gupta and Hero~\cite{gupta2003high}. 
  In this case, function $F(y)$ reduces to:
\[
F(y) = \nabla\!\Lambda(y)\T\,M(y)\,\nabla\!\Lambda(y) \ ,
\]
where $\Lambda(y) = \log\frac{p_0(y)}{p_1(y)}$ is the single sample LLR.
Then, expression~(\ref{eq:loss}) of $D_e$ is consistent with the results of Gupta and Hero (see in particular~\cite[Equation~(20)]{gupta2003high}).

Note that we assume that each joint density $p_0(y_{1:n})$ and $p_1(y_{1:n})$ is of class $C_3$ on $\sY^n$. Gupta and Hero's assumption is weaker, since they only assume that ``\emph{the densities are twice continuously differentiable on an open set of probability $1$}''~\cite[page 1956]{gupta2003high}. In fact, we need conditions on the third derivatives of the logarithm of the densities in order to find relevant upper bounds of the Taylor-Lagrange remainders in the expansion of the \emph{joint} densities $p_i(y_{-m:u})$ in the general case (see the detailed proof in Section~\ref{sec:proof}).
\end{remark}

\begin{remark}
We now provide some insights on the meaning of Assumption~\ref{ass:loss}
and on the class of stationary processes which satisfy the latter.
Assumptions~\ref{ass:loss}-1) and~\ref{ass:loss}-2) are mild technical conditions
on the smoothness of the pdf of the observations. They encompass a large family
of stochastic processes and are generally simple to validate on a case-by-case basis.
As explained above, Assumption~\ref{ass:loss}-3) can be interpreted as a condition
on the speed at which past observations are forgotten. Quantities
$\eta_i(m)$ and $\eta\iN(m)$ can be interpreted as conditional mixing
coefficients associated with the unquantized and quantized processes $(Y_k)_k$ and $(Z\Nk)_k$
respectively (see Remark~\ref{rem:mixing} below). 
Past observations must be forgotten at least at a polynomial speed faster than $m^6$.
Assumption~\ref{ass:loss}-4) can be interpreted similarly as a forgetting property,
which no longer involves the logarithm of the density of the observations, but its derivative.
For instance, Assumption~\ref{ass:loss} is simple to verify in case of short-dependent processes
(such as moving average processes for instance) provided that the density of the observation is smooth enough.
A similar remark holds for a wide class of Markov chains. In this case, Assumption~\ref{ass:loss} essentially reduces
to a smoothness assumption on the density of the transition kernel.
More generally, we prove in Section~\ref{sec:illus} that Assumption~\ref{ass:loss} holds
for a wide class of hidden Markov models: We provide sufficient conditions on the transition
kernel such that Assumption~\ref{ass:loss} holds. See also the numerical results in 
Section~\ref{sec:numerical}.
\end{remark}

\begin{remark}
\label{rem:mixing}
It is worth making some remarks on the link between
  Assumption~\ref{ass:loss} and standard mixing conditions used in the
  literature on mixing
  processes~\cite{doukhan1994mixing,bosq1998nonparametric,bradley2005basic}.
The mixing property which is the closest to our setting is related to the notion of $\psi$-mixing.
For two $\sigma$-fields $\cU$ and $\cV$, define the following coefficient~\cite{doukhan1994mixing,bradley2005basic}:
\[
\psi(\cU,\cV) = \sup_{\substack{U\in\cU,V\in\cV \\ \bP(U)>0,\bP(V)>0}} \abs{ 1-\frac{\bP(U\cap V)}{\bP(U)\,\bP(V)} } \ .
\]
Recall that a stochastic process $Y_\bZ$ is said to be $\psi$-mixing when the sequence of $\psi$-mixing coefficients
$\psi(\sigma(Y_{n+1}) , \sigma(Y_{-\infty:0}))$ converges to zero. The classical $\psi$-mixing condition traduces the fact
that, loosely speaking, current samples at time $n$ tend to become independent of past samples $Y_0, Y_{-1},\dots$
as $n$ increases. In our case, we need to ensure that current samples become independent of past ones
\emph{conditionally to intermediate values} $Y_{1:n}$. Usual $\psi$-mixing coefficient do not
fully allow to grasp this property. In~\cite{villardISIT}, we introduced  the following \emph{conditional}
$\psi$-mixing coefficient for $\sigma$-fields $\cU$, $\cV$ and $\cW$:
\[
   \bar\psi_i(\cU,\cV|\cW) = \sup_{U\in\cU,\,V\in\cV}\ess\sup \abs{ 1- \frac{\bP_i(U\cap V | \cW)}{\bP_i(U| \cW)\,\bP_i(V | \cW)} } 
\]
where the essential supremum is taken w.r.t. $\bP_0$ and where we use
the convention $0/0=1$. The above coefficient can be interpreted as a
measure of dependence (under $\bP_i$) between $\cU$ and $\cV$
conditionally to $\cW$. In particular, it coincides with the
traditional $\psi$-mixing coefficient $\psi(\cU,\cV)$ when
$\cW$ is taken to be the whole space $B(\sY^\bZ)$ and $\bP=\bP_0$. For each $n\geq 1$, we
further define $\bar\psi_i(n) = \bar\psi_i(\sigma(Y_{n+1}),\sigma(Y_{-\infty:0})|\sigma(Y_{1:n}))$
and $\bar\psi_i(0) = \bar\psi_i(\sigma(Y_{1}),\sigma(Y_{-\infty:0}))$ when $n=0$. 
There exists a close link between the above conditional mixing coefficients
and the set of coefficients $\eta_i(m)$ defined in~(\ref{eq:eta}).
In particular, Theorem~2 is valid when Assumption~\ref{ass:loss}-2) is replaced by the assumption
that sequences $\bar\psi_{1}(n)$ and $\bar\psi_{i,N}(n) = \bar\psi_i(\sigma(Z_{N,n+1}),\sigma(Z_{N,-\infty:0})|\sigma(Z_{N,1:n}))$
 converge to zero at speed $n^{6+\epsilon}$. We refer to~\cite{villardISIT} for details.
\end{remark}

The asymptotic loss in error exponent $D_e$ depends on the quantizer 
through its model point density $\zeta$ and its model covariation profile $M$.
In the sequel, we study the values of these parameters which
attenuate as much as possible the loss $D_e$.

\subsection{Determination of Relevant High-Rate Quantizers: Scalar case ($d=1$)}
\label{sec:scalar}

We first address the case where measurements $(Y_k)_{k\geq0}$ are
real-valued.  Assume without much loss of generality that each cell is
connected (cells are intervals) \emph{i.e.}, the quantizer is regular~\cite{gersho1992vector}.
In this case, a straightforward
derivation leads to $M_N(y)= 1/12$ for each $y$ and each $N$.
Therefore, function $F$ simplifies~to:
\begin{eqnarray*}
F(y) &=& \frac1{12}\,\bE_0 \left[\ell(Y_\bZ)^2\,\Big|\,Y_0=y \right]\\
		&=& \frac1{12}\,\lim_{k\to \infty}	\bE_0 \left[\left( \dron{}{y_0}\log\frac{p_0}{p_1}(Y_{-k:k})\right)^2 \,\bigg|\,Y_0=y \right] .
\end{eqnarray*}
Using Holder's inequality on~(\ref{eq:loss}), it is straightforward to prove the
following result.
\begin{coro}
Assume that $d=1$ and that cells are intervals. The error exponent loss $D_e$ is such that:
\begin{equation}
\label{eq:holderscalar}
D_e \geq \frac12\,\left( \int \left[p_0(y)F(y)\right]^{1/3} dy \right)^3 \ ,
\end{equation}
where equality holds in~(\ref{eq:holderscalar}) when the model point density coincides with:
\[
  \zeta(y) = \frac{\left[p_0(y)F(y)\right]^{1/3}}{\int \left[p_0(s)F(s)\right]^{1/3} ds}\ .
\]
\end{coro}
The above corollary provides the optimal high-rate quantization rule
for the initial hypothesis testing problem.
Note that expression~\eqref{eq:holderscalar} is quite similar to~\cite[Equation (15)]{panter1951quantization} which gives ``the minimum distortion resulting with optimum level spacing'' in an MSE perspective.

\begin{remark}
\label{rem:compander}
In practice, $N$-point scalar quantizer achieving a given model point density $\zeta$ can be easily implemented by means of a compander.
Recall that a compander is defined as the composition of an invertible
continuous function $\phi$ (the so-called compressor) and a uniform
quantizer~\cite{bennett1948spectra,gray1998quantization}. To that end,
it is sufficient to define the compressor $\phi$ as the primitive of $\zeta$
on the observation space. For example, if $\sY$ is the segment $[a,b]\subset
\bR$, define $\phi(x)=\int_a^x \zeta(t)dt$. Next the output of the compander is
quantized using a uniform $N$-point quantizer on the interval $[0,1]$.
Under the assumption that $\zeta$ is a Lipschitz function, it is
straightforward to show that the resulting sequence of quantizers
satisfies Assumption~\ref{ass:high-rate} \emph{i.e.}, that it achieves
the model point density $\zeta$.
\end{remark}

\subsection{Determination of Relevant High-Rate Quantizers: Vector case ($d\geq 2$)}

In the vector case, the determination of optimal high-rate quantization rules 
implies the joint minimization of expression~\eqref{eq:loss} w.r.t. both functions $\zeta$ and $M$.
Unfortunately, as remarked in~\cite{neuhoff1996asymptotic,gray1998quantization},
it is not known what functions $M$ are allowable
as covariation profiles. The determination of the set of admissible couples $(\zeta,M)$ is 
an open problem, which is beyond the scope of this paper.

However, when $M$ is fixed, the point density $\zeta$ which minimizes $D_e$ 
can be easily expressed as a function of $M$. Once again, this is a consequence
of Holder's inequality:
\[
D_e \geq \frac12\,\left( \int \left[p_0(y)F(y)\right]^\frac d{d+2} dy \right)^\frac{d+2}d \ ,
\]
where equality is achieved when the point density coincides with:
\begin{equation}
  \label{eq:vector_zetaopt}
  \zeta(y) = \frac{\left[p_0(y)F(y)\right]^\frac d{d+2}}{\int \left[p_0(s)F(s)\right]^\frac d{d+2} ds}\ .
\end{equation}
In other words, one can easily provide the optimal high-rate quantization rule
for a given limiting covariation profile. 
Following the approach of~\cite{gupta2003high}, we study two special cases of covariation profile:

\subsubsection{Congruent cells with minimum moment of inertia}
\label{sec:vector_mini}

In this paragraph, we focus on congruent cells with minimum moment of inertia \emph{i.e.}, we assume that
\begin{equation}
\label{eq:vector_mini}
\forall\,y\in\sY,\ M(y)=\nu I_d\ , 
\end{equation}
for some $\nu>0$, where $I_d$ represents the $d×  d$ identity matrix. 

Recall that Gersho~\cite{gersho1979asymptotically} made the now widely accepted conjecture that when~$N$ tends to infinity, most cells (\emph{i.e.}, all the cells except those which are close to the boundary of the considered domain) of a $d$-dimensional MSE-optimal quantizer become congruent to some tessellating $d$-dimensional polytope $H_d^*$. In such a case, $M(y)$ is independent of $y$. 
Furthermore, Zamir and Feder~\cite[Lemma 1]{zamir1996lattice} proved that the cells of the MSE-optimal lattice quantizers become ``closer'' to balls \emph{i.e.}, with minimum moment of inertia, as dimension $d$ grows.

For quantizers with covariation profile given by~\eqref{eq:vector_mini}, the optimal point density~\eqref{eq:vector_zetaopt} becomes:
\begin{equation}
\label{eq:LBG_zetaopt}
\zeta(y) = \frac{\left[p_0(y)\bar F(y)\right]^\frac d{d+2}}{\int \left[p_0(s)\bar F(s)\right]^\frac d{d+2} ds}\ ,
\end{equation}
where function $\bar F$ is defined by
\begin{eqnarray}
\bar F(y) 
	&=& \bE_0 \left[\norm{\ell(Y_\bZ)}^2\,\Big|\,Y_0=y \right] \nonumber \\
	&=& \lim_{k\to \infty}\bE_0 \left[\left\|\nabl{y_0}\log\frac{p_0}{p_1}(Y_{-k:k})\right\|^2\,\Big|\,Y_0=y \right] . \label{eq:Fbarlim}
\end{eqnarray}

\paragraph*{Design Algorithm}

In practice, one would like to design an $N$-point quantizer which point density approximately equals~\eqref{eq:LBG_zetaopt} for some finite $N$.
This can be achieved by means of well-established algorithms, the most popular of them being the Linde-Buzo-Gray (LBG) algorithm~\cite{linde1980algorithm}. This algorithm is an iterative method which computes a Voronoi tessellation, and yields an MSE-optimal $N$-point quantizer, from a training set of data of some pdf $p_0(y)$.

An ($N$-point) MSE-optimal quantizer for density $p_0(y)$ minimizes $\bE_0\left[\norm{Y_0-\xi_N(Y_0)}^2\right]$. 
As the number of quantization points~$N$ tends to infinity, such a quantizer has the following model point density~\cite{gray1998quantization,na1995bennett}:
\begin{equation}
\label{eq:bennett}
\zeta_{MSE}(y) = \frac{p_0(y)^\frac d{d+2}}{\int p_0(s)^\frac d{d+2}\,ds} \ .
\end{equation}
Comparing Equations~(\ref{eq:LBG_zetaopt}) and~(\ref{eq:bennett}), we deduce that the proposed quantizer, whose model point density $\zeta$ is given by Equation~\eqref{eq:LBG_zetaopt}, can be obtained in practice by simply feeding the classical LBG algorithm with a training set of data of the following pdf:
\[
q^*(y) = \frac{p_0(y){\bar F}(y)}{\int p_0(s){\bar F}(s)\,ds}\ .
\]
Section~\ref{sec:numerical} provides numerical illustrations of this approach.

\subsubsection{Ellipsoidal cells}

In order to yield some insights on the general shape of the cells, and following~\cite{gupta2003high}, we focus in this paragraph on ellipsoidal cells. This kind of cells can not partition the considered convex subset $\sY$ of $\bR^d$ but, for large dimension $d$, in analogy with the spherical cell approximation~\cite{zamir1996lattice,gray1998quantization,conway1999sphere}, we may assume that almost all cells of a given quantizer are close to ellipsoids. 

Such an ellipsoidal cell, in the neighborhood of point $y$ writes $C = \{ x : (x-y)\T R(y)\,(x-y)  \leq 1 \}$, for some symmetric positive definite matrix $R(y)$. The corresponding covariation profile writes $M(y) = \nu \abs{R(y)}^{1/d} R(y)^{-1}$~\cite{gupta2003high,gupta2001quantization}, for some $\nu>0$, and has an eigendecomposition 
\[
M(y) = U(y)\,\Phi(y)\,U(y)\T \ ,
\]
where $\Phi(y) = \diag{\phi_{1:d}(y)}$,\footnote{For any given $d$-dimensional vector $x_{1:d}\in\bR^d$, $\diag{x_{1:d}}$ represents the $d$-by-$d$ diagonal matrix with diagonal coefficients $(x_1,x_2,\dots,x_d)$.} and $U(y)$ is an orthogonal matrix. Note that the (positive) eigenvalues $(\phi_i(y))_{i\in\{1,\dots,d\}}$ of $M(y)$ capture the relative importance of the axes of the ellipsoid~$C$, while the columns $(u_i(y))_{i\in\{1,\dots,d\}}$ of $U(y)$ \emph{i.e.}, the eigenvectors of $M(y)$, indicate their respective directions.

In this paragraph, we assume that eigenvalues $(\phi_i)_{i\in\{1,\dots,d\}}$ are fixed, constant w.r.t. $y$ and, without loss of generality, sorted in increasing order \emph{i.e.}, $0<\phi_1\leq\phi_2\leq\dots\leq\phi_d$. We want to find the best orthogonal matrix $U(y)$ \emph{i.e.}, the one which minimizes function $F(y)$, given at Equation~\eqref{eq:F}, in order to minimize the error exponent loss $D_e$~\eqref{eq:loss}. In other words, for a given ``shape'' of (non-degenerate) ellipsoid, we look for the best directions of its axes.
Function $F(y)$ writes:
\begin{eqnarray}
F(y)&=& \bE_0 \left[\ell(Y_\bZ)\T M(Y_0)\, \ell(Y_\bZ)\,\Big|\,Y_0=y \right] \nonumber\\
	&=& \tr{ U(y)\,\Phi\,U(y)\T \bar L(y) }  \label{eq:ellips} \ ,
\end{eqnarray}
where $\bar L(y)=\bE_0 \left[\ell(Y_\bZ)\,\ell(Y_\bZ)\T \,\Big|\,Y_0=y \right]$.
Now write the eigendecomposition of the positive definite matrix $\bar L(y)$ :
\[
\bar L(y) = V(y)\,\Delta(y)\,V(y)\T \ ,
\]
where $\Delta(y) = \diag{\lambda_{1:d}(y)}$, $(\lambda_i(y))_{i\in\{1,\dots,d\}}$ are the (positive) eigenvalues of $\bar L(y)$ sorted in increasing order \emph{i.e.}, $0<\lambda_1(y)\leq\lambda_2(y)\leq\dots\leq\lambda_d(y)$, and $V(y)$ is an orthogonal matrix.
Equation~\eqref{eq:ellips} thus writes:
\begin{eqnarray*}
F(y)&=& 	\tr{ U(y)\,\Phi\,U(y)\T V(y)\,\Delta(y)\,V(y)\T } \\
	&\geq&	\sum_{i=1}^d 	\phi_i\,\lambda_{d-i+1}(y) \ ,
\end{eqnarray*}
where the last inequality follows from a well-known trace inequality for positive semidefinite Hermitian matrices~\cite{lasserre1995trace},~\cite[Section 9-H]{marshall1979inequalities}. The above lower bound is furthermore achieved choosing matrix $U(y)$ such that $U(y)\T V(y)$ is the anti-diagonal matrix with ones on its anti-diagonal \emph{i.e.}, defining the $i$th column of matrix $U(y)$ as the $(d-i+1)$th column of matrix $V(y)$, or equivalently eigenvector $u_i(y)$ of matrix $M(y)$ as eigenvector $v_{d-i+1}(y)$ of matrix $\bar L(y)$.

From the above derivations, we conclude that if a cell is a non-degenerate ellipsoid around $y$ then its axes should be aligned along the ones of matrix $\bar L(y)$ in the reverse order. In particular,  its minor axis should be aligned along the principal eigenvector of matrix $\bar L(y)$.

\section{Proof of Theorem~\ref{th:cv_ordre2}}
\label{sec:proof}

\subsection{Preliminaries}

Recall that $V\Nj = \int_{C\Nj} dy$ is the volume of cell $C\Nj$ ($j\in\{1,\dots,N\}$).
For each $i\in\{0,1\}$ and each set of quantization points
$\xi_{N,j_{1:n}}=(\xi_{N,j_1},\dots,\xi_{N,j_n})\in\Xi_N^n$,
define the following rescaled pdf of $Z_{N,1:n}$:
\begin{eqnarray}
\bar p\iN(\xi_{N,j_{1:n}})
&=& \frac1{V_{N,j_1}× \dots × V_{N,j_n}}\ p\iN(\xi_{N,j_{1:n}})\nonumber \\
&=& \frac1{V_{N,j_1}× \dots × V_{N,j_n}}\ P_{i,n}(C_{N,j_1}× \dots× C_{N,j_n})\ .
\label{eq:def_puN}
\end{eqnarray}
The above definition is convenient because 
$\bar p\iN(\xi_{N,j_{1:n}}) \simeq   p_{i}(\xi_{N,j_{1:n}})$
for large values of $N$. This approximation will be of prime importance in the sequel.
We define function $\bar p\iN(\xi_{N,j_n}|\xi_{N,j_{1:n-1}})$ similarly.

For each $i\in\{0,1\}$ and each integer $\ell\geq0$,
we introduce the following functions:
\begin{align*}
&\forall\ y_{-\ell:0}\in \sY^{\ell+1},\quad
	\cL_i(y_{-\ell:0}) = \log p_i(y_0 | y_{-\ell:-1}) \ ,\\
&\forall\ z_{-\ell:0}\in \Xi_N^{\ell+1}, \quad
	\cL\iN(z_{-\ell:0}) = \log \bar p\iN(z_0 | z_{-\ell:-1}) \ .
\end{align*}

Due to Assumptions~\ref{ass:dens}-3) and~\ref{ass:loss}-3)
(which ensures that $\eta_i(0)<\infty$), random sequence
$(\cL_i(Y_{-\ell:0}))_{\ell\geq0}$ lies in $L^1(P_0)$.
Moreover, Assumption~\ref{ass:loss}-3) for large~$m$
ensures that sequence $(\cL_i(Y_{-\ell:0}))_{\ell\geq0}$
is a Cauchy sequence of~$L^1(P_0)$.
Denote by $\cL_i(Y_{-\infty:0})$ its limit.
From Assumption~\ref{ass:loss}-3) once again, the following holds for any $\ell\geq0$,
\begin{equation}
\label{eq:oubli6}
\bE_0 |\cL_i(Y_{-\ell:0}) - \cL_i(Y_{-\infty:0})| \leq \frac{C_e}{(1+\ell)^{\,6+\epsilon}} \ .
\end{equation}

A similar result holds for sequence $(\cL\iN(Z_{N,-\ell:0}))_{\ell\geq0}$ 
which converges in $L^1(\bP_0)$ to some random variable $\cL\iN(Z_{N,-\infty:0})$
and verifies for any $\ell\geq0$,
\begin{equation}
\label{eq:oubliN6}
\bE_0 |\cL\iN(Z_{N,-\ell:0}) - \cL\iN(Z_{N,-\infty:0})| \leq \frac{C_e}{(1+\ell)^{\,6+\epsilon}}\ . 
\end{equation}

Our aim is to study the difference $K-K_N$ between error exponents associated
with the ideal and quantized cases respectively. We may write the difference as
\begin{equation}
\label{eq:dec_K-KN}
K-K_N = (K_0-K\oN)-(K_1-K\uN) \ ,
\end{equation}
where we defined for each $i\in\{0,1\}$,
\begin{eqnarray*}
K_i  &=& \bE_0\left[\cL_i(Y_{-\infty:0})\right] \ ,\\
K\iN &=& \bE_0\left[\cL\iN(Z_{N,-\infty:0})\right]\ . 
\end{eqnarray*}
In the sequel, we focus on the study of $K_1-K\uN$, the study of $K_0-K\oN$
being similar. 

We now proceed with the proof.  Choose any $\epsilon'$ such that $0< \epsilon' <
\frac{\epsilon}{3d(6+\epsilon)}$.  Define the sequence of integers $m=m(N)=\lfloor N^{1/(3d)-\epsilon'}\rfloor$.
We shall remember that with this definition,
\begin{equation}
 \lim_{N\to\infty} \frac {m^3}{N^{1/d}} = 0 \ . 	\label{eq:mN_zero}
\end{equation}
The following decomposition holds true:
$K\uN = K_1 + T_N + U_N +\delta_N$, where we defined:
\begin{align*}
  T_N 			&= \bE_0\left[	\cL\uN(Z_{N,-m:0}) - \cL_1(Z_{N,-m:0}) 	\right] \ ,\\
  U_N 			&= \bE_0\left[	\cL_1(Z_{N,-m:0}) 	- \cL_1(Y_{-m:0}) 		\right] \ ,\\
  \delta_N &= \bE_0\left[ \cL\uN(Z_{N,-\infty:0}) - \cL\uN(Z_{N,-m:0}) \right]
  +\bE_0\left[ \cL_1(Y_{-m:0}) - \cL_1(Y_{-\infty:0}) \right]\ .
\end{align*}
Using Equations~\eqref{eq:oubli6} and~(\ref{eq:oubliN6}), it is straightforward to show that
\[
  N^{2/d}|\delta_N| \leq 2\,C_e\,\frac{N^{2/d}}{(1+m)^{\,6+\epsilon}}\ .
\]
By definition of $m=m(N)$, we deduce that $N^{2/d}|\delta_N|$ converges to zero
as $N\to\infty$. As a consequence, the asymptotic analysis of quantity
$N^{2/d}(K\uN-K_1)$ reduces to the study of $T_N$ and $U_N$. 

As $\sY$ is a bounded set, Assumption~\ref{ass:loss}-2) implies the following
bounds on the derivatives of density~$p_1$ which will be of permanent use in the sequel:
\begin{eqnarray}
\sup_{\{y_{1:n}\in \sY^{n},\;1\leq k\leq n\}}\ \norm{\nabl{y_k}\log p_1(y_{1:n})} &\leq& C_1 \ , \label{eq:C1}\\
\sup_{\{y_{1:n}\in \sY^{n},\;1\leq k\leq n\}}\ \norm{\hess{y_k}\log p_1(y_{1:n})} &\leq& C_2 \ , \label{eq:C2}
\end{eqnarray}
for some constants $C_1$ and $C_2$.

\subsection{Study of $T_N$}
\label{sec:TN}

We expand $T_N$ as follows: 
\begin{equation}
\label{eq:exp_TN}
T_N = \bE_0\left[ \log \frac{\bar p\uN(Z_{N,-m:0})}{p_1(Z_{N,-m:0})} \right]
	- \bE_0\left[ \log \frac{\bar p\uN(Z_{N,-m:-1})}{p_1(Z_{N,-m:-1})} \right] \ .
\end{equation}
We now study each term of the r.h.s. of the above equality.
Consider $u\in \{-1,0\}$.
Writing the Taylor-Lagrange expansion of function 
$y_{-m:u} \mapsto p_1(y_{-m:u})$ at point $\xi_{N,j_{-m:u}}$,
using Assumptions~\ref{ass:high-rate}-3),~\ref{ass:loss} and the properties of the quantizers sequence,
we prove the following lemma (the detailed proof is given in Appendix~\ref{app:dev_puN_p1}).
\begin{lemma}
\label{lem:dev_puN_p1}
For each $j_{-m:u}\in \{1,\dots,N\}^{u+m+1}$,
the following expansion holds true:
\[
\frac{\bar p\uN(\xi_{N,j_{-m:u}})}{p_1(\xi_{N,j_{-m:u}})} = 1 + \frac1{2
  N^{2/d}} \sum_{k=-m}^u
\tr{\frac{\hess{y_k}p_1(\xi_{N,j_{-m:u}})\T}{p_1(\xi_{N,j_{-m:u}})}\,\frac{M\Njk}{\zeta\Njk^{2/d}}}
+ \epsilon_{N,j_{-m:u}}\ ,
\]
where $\abs{\epsilon_{N,j_{-m:u}}} \leq c_T \left(\frac{m+1}{N^{1/d}}\right)^3$ for some constant $c_T$.
\end{lemma}

Plugging the above equation into~(\ref{eq:exp_TN}), using $\big| \log(1+x) - x \big| \leq x^2$ in a neighborhood of zero, Assumptions~\ref{ass:high-rate},~\ref{ass:loss}-2) and Equation~\eqref{eq:mN_zero}, we obtain:
\begin{equation}
\label{eq:TN}
T_N = T_N(0) - T_N(-1) + o_N(N^{-2/d}) \ ,
\end{equation}
where, for each $u\in\{-1,0\}$,
\begin{equation}
\label{eq:TNu}
T_N(u) = \frac1{2 N^{2/d}} \sum_{k=-m}^u
	\bE_0 \left[ \tr{\frac{\hess{y_k}p_1(Z_{N,-m:u})\T}{p_1(Z_{N,-m:u})}\,\frac{M_N(Y_k)}{\zeta_N(Y_k)^{2/d}}} \right] \ .
\end{equation}

\subsection{Study of $U_N$}
\label{sec:UN}

We expand $U_N$ as follows: 
\begin{equation}
\label{eq:exp_UN}
U_N = \bE_0\left[ \log p_1(Z_{N,-m:0}) - \log p_1(Y_{-m:0}) \right]
	- \bE_0\left[ \log p_1(Z_{N,-m:-1}) - \log p_1(Y_{-m:-1}) \right] \ ,
\end{equation}
and study each term of the r.h.s. of the above equality.
For each $u\in \{-1,0\}$ and each $j_{-m:u}\in \{1,\dots,N\}^{u+m+1}$,
we expand function $y_{-m:u} \mapsto \log p_1(y_{-m:u})$
at point $\xi_{N,j_{-m:u}}$:
\begin{multline}
\label{eq:dev_logp}
\log p_1(y_{-m:u}) = \log p_1(\xi_{N,j_{-m:u}}) 
  + \sum_{k=-m}^u \nabl{y_k}\log p_1(\xi_{N,j_{-m:u}})\T\,(y_k-\xi\Njk) \\
  + \frac{1}{2} \sum_{k,\ell=-m}^u (y_k-\xi\Njk)\T\,\hess{y_k,y_\ell}\log p_1(\xi_{N,j_{-m:u}})\,(y_\ell-\xi\Njl)
  + \epsilon'_N(y_{-m:u}) \ .
\end{multline}
Under Assumptions~\ref{ass:high-rate}-3) and~\ref{ass:loss}-2), for each $y_{-m:u}\in C_{N,j_{-m}}×  \cdots ×  C_{N,j_u}$,
the remainder is such that
\[
\abs{\epsilon'_N(y_{-m:u})} \leq (m+1)^3\,c_3' \left(\frac{C_d}{N^{1/d}}\right)^3 \ ,
\]
for some constant $c_3'$.
By Equation~\eqref{eq:mN_zero}, the r.h.s. of the above inequality converges
to zero as $N$ tends to infinity faster than $N^{-2/d}$.
Plugging Taylor expansion~\eqref{eq:dev_logp} into the expression~\eqref{eq:exp_UN} of $U_N$,
we obtain:
\begin{equation}
\label{eq:UN}
U_N = U_N(0) - U_N(-1) + o_N(N^{-2/d}) \ ,
\end{equation}
where, for each $u\in \{-1,0\}$,
\begin{multline}
\label{eq:def_UNu}
U_N(u) =- \sum_{k=-m}^u \bE_0\left[ \nabl{y_k}\log p_1(Z_{N,-m:u})\T\,(Y_k-Z\Nk) \right] \\
		- \frac{1}{2} \sum_{k,\ell=-m}^u \bE_0\left[(Y_k-Z\Nk)\T\,\hess{y_k,y_\ell}\log p_1(Z_{N,-m:u})\,(Y_\ell-Z\Nl)\right] \ .
\end{multline}

The next step is to study each dominant term of the r.h.s. of~\eqref{eq:def_UNu}.
The proof of the following lemma is provided in Appendix~\ref{app:dev_UNu}.
\begin{lemma}
\label{lem:dev_UNu}
The following equality holds true for each $u\in \{-1,0\}$:
\[
U_N(u) = A_N(u) + B_N(u) + o_N(N^{-2/d})\ ,
\]
where $A_N$ and $B_N$ are defined as follows:
\begin{eqnarray}
\label{eq:ANu}
A_N(u) &=& - \frac1{N^{2/d}} \sum_{k=-m}^u	
			\bE_0\left[ \nabl{y_k}\log p_1(Z_{N,-m:u})\T \frac{M_N(Y_k)}{\zeta_N(Y_k)^{2/d}}\,
			\nabl{y_k}\log p_0(Y_{-m:u}) \right] \ , \\
B_N(u) &=& - \frac1{2 N^{2/d}} \sum_{k=-m}^u
			\bE_0\left[\tr{\hess{y_k}\log p_1(Z_{N,-m:u}) \frac{M_N(Y_k)}{\zeta_N(Y_k)^{2/d}}} \right] \ .
\nonumber
\end{eqnarray}
\end{lemma}

Now we expand the term $\hess{y_k}\log p_1$ as follows:
\[
\hess{y_k}\log p_1(y_{-m:u})
	= \frac{\hess{y_k}p_1(y_{-m:u})}{p_1(y_{-m:u})}
	- \frac{\nabl{y_k}p_1(y_{-m:u})\,\nabl{y_k}p_1(y_{-m:u})\T}{(p_1(y_{-m:u}))^2} \ .
\]

From the above decomposition and Equation~\eqref{eq:TNu},
we can divide $B_N(u)$ into two terms:
\[
B_N(u) = \frac1{2 N^{2/d}} \sum_{k=-m}^u
			\bE_0\left[\tr{ \nabl{y_k}\log p_1(Z_{N,-m:u})\,\nabl{y_k}\log p_1(Z_{N,-m:u})\T
			\frac{M_N(Y_k)}{\zeta_N(Y_k)^{2/d}}} \right] - T_N(u) \ .
\]

Expanding function $\nabl{y_k}\log p_1$ in the above equation and in~\eqref{eq:ANu},
we can write dominant terms in a simple form \emph{i.e.}, replace each $Z_N$ by $Y$.
Under Assumption~\ref{ass:high-rate}, from Equations~\eqref{eq:C2} and~\eqref{eq:mN_zero},
we can easily prove that the corresponding remainders are $o_N(N^{-2/d})$.

Putting all pieces together, we obtain
\begin{multline}
\label{eq:UNu_fin}
U_N(u) =- \frac1{N^{2/d}} \sum_{k=-m}^u	
			\bE_0\left[ \nabl{y_k}\log p_1(Y_{-m:u})\T \frac{M_N(Y_k)}{\zeta_N(Y_k)^{2/d}}\,
			\nabl{y_k}\log p_0(Y_{-m:u}) \right] \\
		+ \frac1{2 N^{2/d}} \sum_{k=-m}^u
			\bE_0\left[ \nabl{y_k}\log p_1(Y_{-m:u})\T \frac{M_N(Y_k)}{\zeta_N(Y_k)^{2/d}}\,
			\nabl{y_k}\log p_1(Y_{-m:u}) \right] \\
		- T_N(u) +o_N(N^{-2/d}) \ .
\end{multline}

\subsection{End of the Proof}
\label{sec:end_proof}

From the results of sections~\ref{sec:TN} and~\ref{sec:UN}, 
we can easily prove the following lemma.
\begin{lemma}
\label{lem:K_KN}
The following holds true:
\begin{equation}
\label{eq:K-KN}
N^{2/d} (K-K_N) 
		= \bE_0\left[ \cH_{N,0}(Y_{-m:0}) \right] 
		+ \sum_{k=-m}^{-1} \bE_0\left[\cH\Nk(Y_{-m:0}) - \cH\Nk(Y_{-m:-1}) \right] + o_N(1) \ ,
\end{equation}
where for each $u\in\{-1,0\}$, each $m\geq 1$ and each $k\in\{-m,\dots,u\}$:
\begin{equation}
\label{eq:def_HNk}
\cH\Nk(Y_{-m:u}) = 	\frac12\ \nabl{y_k}\log\frac{p_0}{p_1}(Y_{-m:u})\T\,
					\frac{M_N(Y_k)}{\zeta_N(Y_k)^{2/d}}\,
					\nabl{y_k}\log\frac{p_0}{p_1}(Y_{-m:u}) \ .
\end{equation}
\end{lemma}

\begin{IEEEproof}
Recalling the decomposition: $K\uN = K_1 + T_N + U_N + o_N(N^{-2/d})$
and gathering Equations~\eqref{eq:TN}, \eqref{eq:UN}, \eqref{eq:UNu_fin},
it is straightforward to prove the following
equality:
\begin{align*}
N^{2/d} (K\uN - K_1) =
	&- \sum_{k=-m}^0	
			\bE_0\left[ \nabl{y_k}\log p_1(Y_{-m:0})\T\,\frac{M_N(Y_k)}{\zeta_N(Y_k)^{2/d}}\,
			\nabl{y_k}\log p_0(Y_{-m:0}) \right] \\
	&\hspace{1.5cm}	+ \frac12 \sum_{k=-m}^0
			\bE_0\left[ \nabl{y_k}\log p_1(Y_{-m:0})\T\,\frac{M_N(Y_k)}{\zeta_N(Y_k)^{2/d}}\,
			\nabl{y_k}\log p_1(Y_{-m:0}) \right] \\
	&+ \sum_{k=-m}^{-1}	
			\bE_0\left[ \nabl{y_k}\log p_1(Y_{-m:-1})\T\,\frac{M_N(Y_k)}{\zeta_N(Y_k)^{2/d}}\,
			\nabl{y_k}\log p_0(Y_{-m:-1}) \right] \\
	&\hspace{1.5cm}	- \frac12 \sum_{k=-m}^{-1}
			\bE_0\left[ \nabl{y_k}\log p_1(Y_{-m:-1})\T\,\frac{M_N(Y_k)}{\zeta_N(Y_k)^{2/d}}\,
			\nabl{y_k}\log p_1(Y_{-m:-1}) \right] \\
	& 	+ o_N(1) \ .
\end{align*}

Similar expression holds for $N^{2/d} (K\oN - K_0)$--replace all $p_1$ by $p_0$ in the above equation.
Lemma~\ref{lem:K_KN} follows from decomposition~\eqref{eq:dec_K-KN}.
\end{IEEEproof}

We now study the series~\eqref{eq:K-KN}.
From Assumptions~\ref{ass:high-rate},~\ref{ass:loss}-2) and~\ref{ass:loss}-4), 
the following forgetting properties hold true
for any positive integers $\ell'$, $\ell$ and any integers $k$, $u$ s.t. $-\ell'\leq-\ell\leq k\leq u$:
\begin{eqnarray}
\bE_0 \abs{ \cH\Nk(Y_{-\ell:u}) - \cH\Nk(Y_{-\ell':u})}
	&\leq& c_h \varphi_{\ell-\abs{k}} \ , 	\label{eq:nopast}\\
\bE_0 \abs{\cH\Nk(Y_{-\ell:0}) - \cH\Nk(Y_{-\ell:-1})}
	&\leq& c_h \psi_{\abs{k}} \ , 			\label{eq:nofuture}
\end{eqnarray}
for some constant $c_h$.

It is clear from~\eqref{eq:nopast} that sequence
$\left(\cH\Nk(Y_{-\ell:u}))\right)_{\ell\geq -u}$
is a Cauchy sequence in $L^1(\bP_0)$.
We simply denote its limit by $\cH\Nk(Y_{-\infty:u})$.
Inequalities~\eqref{eq:nopast} and~\eqref{eq:nofuture} provide the main tools
for the asymptotic analysis of series~\eqref{eq:K-KN}.
The proof of the following lemma is given in Appendix~\ref{app:series}.

\begin{lemma}
\label{lem:series}
The following holds true:
\[
N^{2/d} (K-K_N) 
		=	\bE_0\left[\cH_{N,0}(Y_{-\infty:0}) \right]
		+ \sum_{k=-\infty}^{-1}	\bE_0\left[ \cH\Nk(Y_{-\infty:0}) - \cH\Nk(Y_{-\infty:-1}) \right]
		+ o_N(1) \ .
\]
\end{lemma}

As process $(Y_k)_{k\in\bZ}$ is stationary, the expectation $\bE_0$ enclosed in the sum
of the above equation is invariant w.r.t. a time-shift. Using this remark,
we obtain after algebra
\begin{equation}
\label{eq:K-KN_simple}
N^{2/d} (K-K_N) = \lim_{k\to\infty} \bE_0\left[\cH_{N,0}(Y_{-\infty:k})\right]+ o_N(1) \ .
\end{equation}

For a fixed $k\geq0$, Equation~\eqref{eq:phi} ensures that sequence 
$\left( \nabl{y_0}\log\frac{p_0}{p_1}(Y_{-m:k}) \right)_{m\geq 0}$
is a Cauchy sequence in $L^1(\bP_0)$. Denote its limit by $\ell_k(Y_{-\infty:k})$.
The upper bound of Equation~\eqref{eq:psi} is uniform in $m$. Consequently,
it also holds for sequence $\left( \ell_k(Y_{-\infty:k}) \right)_{k\geq0}$:
\[
\bE_0 \norm{ \ell_k(Y_{-\infty:k}) - \ell_{k-1}(Y_{-\infty:k-1}) } \leq \psi_k \ .
\]

Under Assumption~\ref{ass:loss}-4), $\sum_k\psi_k$ is a convergent series.
Sequence $\left( \ell_k(Y_{-\infty:k}) \right)_{k\geq0}$ is thus a Cauchy sequence
in $L^1(\bP_0)$. Denote its limit by $\ell(Y_\bZ)$.
Moreover, the upper bound of Equation~\eqref{eq:phi} (resp. Equation~\eqref{eq:psi})
is uniform in $m'$ (resp. $m$). It is then straightforward to prove that 
$\ell(Y_\bZ)$ coincides with the $L^1(\bP_0)$-limit of 
sequence $\left( \nabl{y_0}\log\frac{p_0}{p_1}(Y_{-k:k}) \right)_{k\geq 0}$.

From Equation~\eqref{eq:C1} and its counterpart for density~$p_0$,
quantity $\nabl{y_0}\log\frac{p_0}{p_1}(Y_{-k:k})$ 
is uniformly bounded. 
Consequently, the above limit also holds in the $L^2(\bP_0)$-sense:
\begin{equation}
\label{eq:ell}
\nabl{y_0}\log\frac{p_0}{p_1}(Y_{-k:k}) \xrightarrow[k\to\infty]{L^2(\bP_0)} \ell(Y_\bZ)\ .
\end{equation}

Plugging Equations~\eqref{eq:def_HNk} and~\eqref{eq:ell} in Equation~\eqref{eq:K-KN_simple}
and letting $N$ tend to $\infty$ complete the proof of Theorem~\ref{th:cv_ordre2}.

\section{Illustration: Case of a Hidden Markov Process}
\label{sec:illus}

In this section, we translate our assumptions in the case of (discrete-time) hidden Markov models. For such models, they reduce to simpler conditions on the transition kernel of the underlying Markov chain, and on the observation kernel.
This context, where the measurements are noisy samples of a certain Markov source, has raised a deep interest in the recent literature on sensor networks (see \cite{hachem2009error,sung2006neyman} and reference therein).

Consider a stationary Markov process $(X_k)_{k\geq 0}$ taking its values in an arbitrary state space $\sX$, and playing the role of a source signal to be detected. 
For each $i\in\{0,1\}$ and each integer $t$, we assume that the (iterated) transition kernel $\bP_i\left[X_{k+t}\in \cdot \,|\, X_k=x\right]$ admits a density $x'\mapsto q_i^t(x,x')$ w.r.t. some probability measure $\lambda$ on $(\sX,B(\sX))$. 
Assume that there exist an integer $m$, and two real numbers $\sigma^-$, $\sigma^+$ s.t., for each $i\in\{0,1\}$ and each $(x,x')\in \sX^2$, $0<\sigma^-\leq q_i^m(x,x')\leq\sigma^+$.
In particular, this assumption implies that the Markov chain $(X_k)_{k\in\bZ}$ has bounded support. 

If the state space $\sX$ is finite, the above conditions hold if the Markov chain $(X_k)_{k\in\bZ}$ is \emph{irreducible aperiodic}, choosing $\lambda$ as the (normalized) counting measure on $\sX$. In this case, the chain indeed admits a stationary distribution, and $q_i^m(x,x')>0$ for each $x$, $x'$ and some integer $m$~\cite[Section~8]{billingsley1995probability}.

\begin{figure}
\centering
\includegraphics[width=9cm]{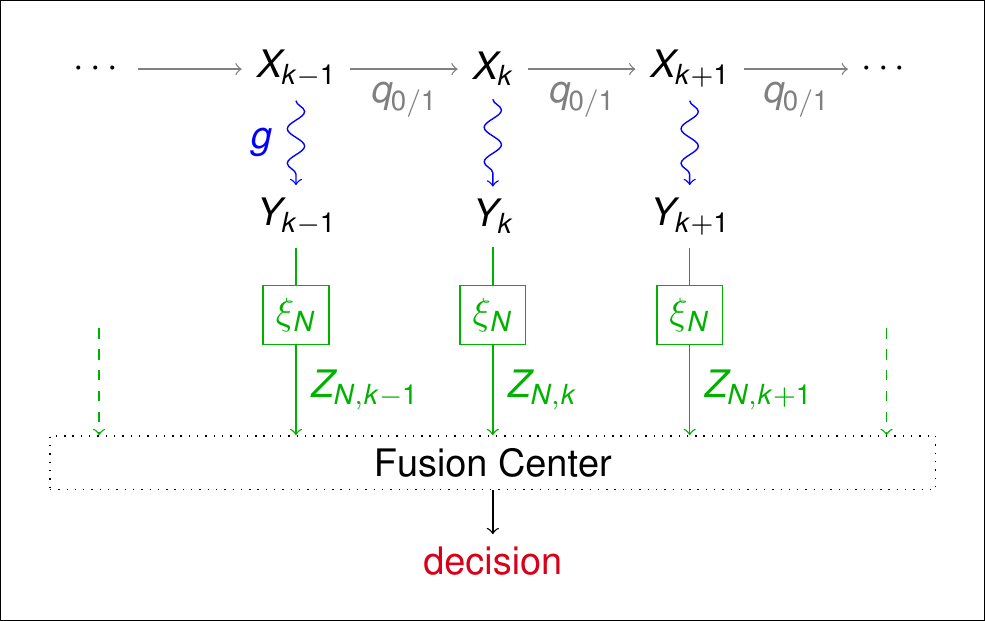}
\caption{Detection of a discrete-time Markov process based on noisy observations.}
\label{fig:hmm}
\end{figure}

The states $X_k$ of the above Markov source are supposed to be hidden.
However, a ``noisy'' version $Y_k$ ($\in\sY\subset\bR^d$) of $X_k$ is available at the $k$th sensor. 
We assume that the distribution $\bP[Y_k\in\cdot\,|X_k=x]$ does not depend on the hypothesis $\HO$ or $\HU$, and 
admits a density $y\mapsto g(x,y)$ w.r.t. the $d$-dimensional Lebesgue measure $\mu$ restricted to $\sY$, such that $0<\inf_{x,y} g(x,y)\leq\sup_{x,y} g(x,y)<\infty$. We furthermore assume that this density verifies some smoothness conditions:
For each $x\in \sX$, $y\mapsto g(x,y)$ is of class $C_3$ on $\sY$, and $\sup_{\{x\in\sX,\; y\in\sY,\; 1\leq h,\bar\i,\bar\jmath\leq d\}} \linebreak	\abs{\dtron{g}{y^{(h)}}{y^{(\bar\i)}}{y^{(\bar\jmath)}}(x,y)}  <\infty$.
The situation is depicted in Figure~\ref{fig:hmm}.

A similar assumption was recently introduced by~\cite{douc2004asymptotic,cappe2007inference} in order to study the asymptotic behaviour of the log-likelihood $\log p_i(Y_{1:n})$ as $n$ tends to infinity. In particular, it was shown that:
\[
\left|\log p_i(Y_0|Y_{-m:-1})-\log p_i(Y_0|Y_{-m':-1})\right| 
	\leq \frac 2{1-\sigma^-/\sigma^+} \left(\frac{\sigma^-}{\sigma^+}\right)^{m-1}
\]
for each $m'\geq m\geq 0$. 
This clearly proves that sequence $\log p_i(Y_0|Y_{-m:-1})$ converges in $L^1(\bP_0)$ as $m\to\infty$ and yields Assumption~\ref{ass:logp}. 
Moreover, the convergence holds at exponential speed, meaning that quantities $\eta_i(m)$, defined by Equation~\eqref{eq:eta}, vanish faster than $1/m^6$.
The same claim holds as well for quantities $\eta\iN(m)$, without need for any special condition on the quantizer (quantization preserves the hidden Markov nature of the original process $(Y_k)_{k\in\bZ}$).
This yields Assumption~\ref{ass:loss}-3).

Assumptions~\ref{ass:loss}-1) and ~\ref{ass:loss}-2) are direct consequences of the above smoothness conditions on density $g$. Assumption~\ref{ass:loss}-4) can be derived following the arguments of~\cite{douc2004asymptotic,cappe2007inference}.

The following proposition then follows from the results of~\cite{douc2004asymptotic,cappe2007inference}. The proof is therefore omitted.
\begin{prop}
\label{prop:hmm}
All conditions given by Assumptions~\ref{ass:dens} and~\ref{ass:loss} hold true for the particular process $(Y_k)_{k\in\bZ}$ described in this section.
\end{prop}

As a consequence, if the family of quantizers moreover verifies Assumption~\ref{ass:high-rate}, then the conclusions of Theorems~\ref{th:exp_err} and~\ref{th:cv_ordre2} hold true. 

Section~\ref{sec:modulation} below provides a practical example of such a detection problem.

\section{Numerical Results}
\label{sec:numerical}

In this section, we provide numerical illustrations of the proposed quantization rule
in terms of geometric properties and performance. 
Different contexts are considered and we compare several quantizers:
\begin{itemize}
\item The \emph{proposed quantizer}, obtained using the approach described in Section~\ref{sec:vector_mini} and whose model point density is given by~(\ref{eq:LBG_zetaopt}).

\item The \emph{MSE-optimal quantizer}, which minimizes $\bE_0\left\|Y_0-Z_{N,0}\right\|^2$ and whose model point density is given by~(\ref{eq:bennett}).

\item \emph{Gupta-Hero quantizer}, introduced in~\cite{gupta2003high}: In this case the model point density is drawn as if observations were i.i.d. \emph{i.e.}, only taking the marginal distributions $p_0(y)$ and $p_1(y)$ into account.

\item The \emph{uniform quantizer} with constant model point density.
\end{itemize}

\subsection{Scenario~\#1: Detection of Quaternary Modulations: QPSK vs. OQPSK}
\label{sec:modulation}

In this section, we provide an example of hidden Markov models which verify the assumptions given at Section~\ref{sec:illus}, and detail how to use in this case the approach described in Section~\ref{sec:vector_mini} for the design of practical quantizers.

\subsubsection{Observation Model}

\begin{figure}[!t]
\centering
\begin{tabular}{cc}
\includegraphics[width=7cm]{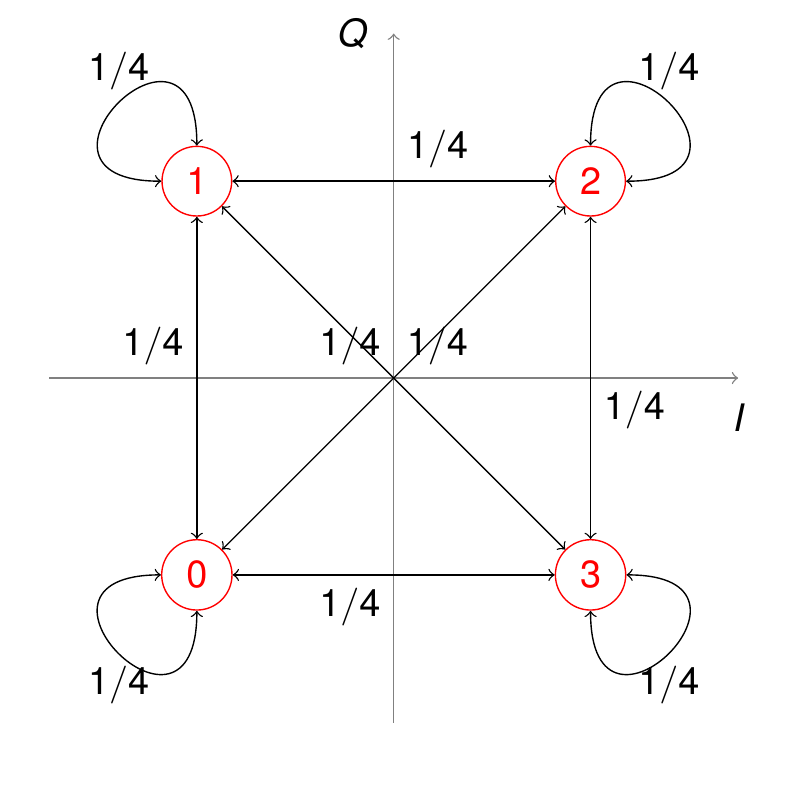}
\quad&\quad
\includegraphics[width=7cm]{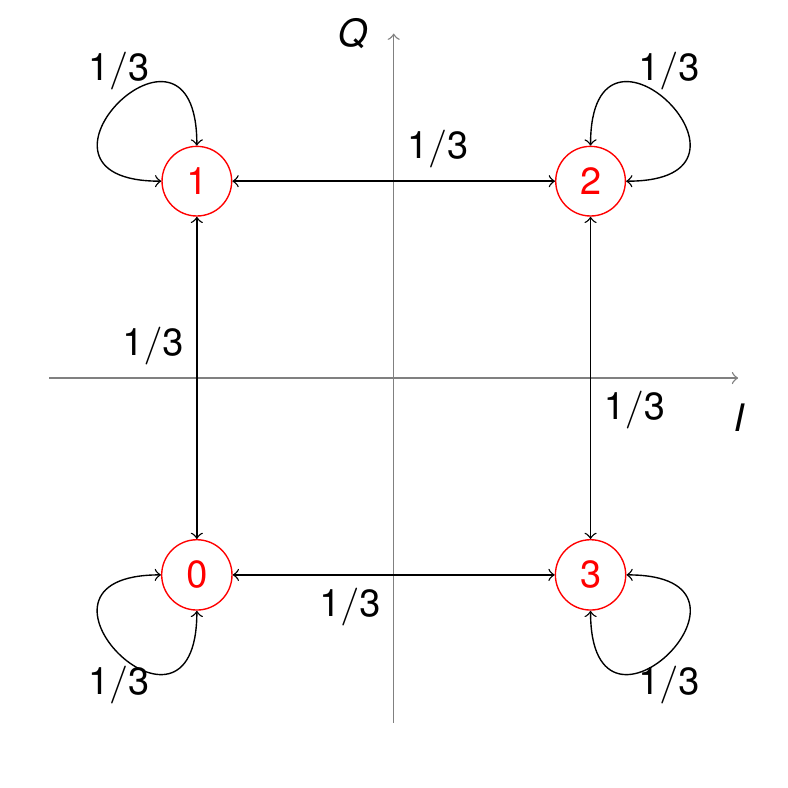}
\\
(a)												& (b)
\end{tabular}
\caption{QPSK vs. OQPSK --
			Constellation diagrams and transitions probabilities for \quad
			(a) QPSK, \quad
			(b) OQPSK.}
\label{fig:QPSKvsOQPSK}
\end{figure}

We consider the following model for vector observations with dimension $d=2$:
\begin{equation}
\label{eq:QPSKvsOQPSK-obs}
Y_k = T(X_k) + W_k \ ,
\end{equation}
where $(X_k)_{k\in\bZ}$ is a 2-bit message, which takes values in $\sX=\{0,1,2,3\}$, 
$T(x)$ is the $2$-D representation of state $x$ in the I-Q plane\footnote{
$	T(0) = [ -1 ; -1 ] ,\
	T(1) = [ -1 ;  1 ] ,\
	T(2) = [  1 ;  1 ] ,\
	T(3) = [  1 ; -1 ]$.
} 
according to Figure~\ref{fig:QPSKvsOQPSK},
and $W_k \stackrel{i.i.d.}{\sim} \cC\cN(0,\sigma^2)$ represents a zero mean  circular Gaussian thermal noise with variance $\sigma^2$. Process $(X_k)_{k\in\bZ}$ is i.i.d., uniformly distributed under $\HO$, and forms a Markov chain under $\HU$. More precisely,
\[
\begin{array}{l}
  \HO :\  X_k \stackrel{i.i.d.}{\sim} \cU_{\{0,1,2,3\}} \\
  \HU :\  X_0\ \sim\ \cU_{\{0,1,2,3\}},\ \bP_1[X_{k+1}=x' | X_k=x]=q(x,x') \ ,
\end{array}
\]
where $q$ is the transition matrix of the Markov chain and is given by:
\[
q = 
\left[
\begin{array}{cccc}
1/3 & 1/3 & 0   & 1/3	\\
1/3 & 1/3 & 1/3 & 0		\\
0   & 1/3 & 1/3 & 1/3	\\
1/3 & 0   & 1/3 & 1/3
\end{array}
\right].
\]
This situation arises when testing from noisy observations between two possible quaternary modulations, namely quadrature phase-shift keying (QPSK) and offset quadrature phase-shift keying (OQPSK), in the In-phase/Quadrature plane~\cite[Chapter~3]{proakis2007digital}. The corresponding constellations are depicted in Figure~\ref{fig:QPSKvsOQPSK}.

In the observation model~\eqref{eq:QPSKvsOQPSK-obs}, densities have infinite support. We thus consider truncated observations on $\sY=[-M;M]^2$ for some positive real number $M$~\cite[Section 10.1]{johnson1994continuous}. The new (truncated) model is a hidden Markov model with observation density $g(x,y)$ given by:
\begin{equation}
\label{eq:QPSKvsOQPSK-g}
g(x,y) = \frac{\ind_{[-M;M]^2}(y)}{C_M(\sigma)} \exp\left(\frac{-1}{2\sigma^2} (y-T(x))\T (y-T(x))\right) \ ,
\end{equation}
where $\ind_A$ stands for the indicator function of set $A$, 
and $C_M(\sigma)$ is a constant such that $\int_{\sY} g(x,y) dy = 1$, for each $x\in\{0,1,2,3\}$ \emph{i.e.}, $C_M(\sigma) = \left( \int_{-M}^M \exp\left(\frac{-(t-1)^2}{2\sigma^2}\right) dt \right)^2$.

The above hidden Markov model verifies the assumptions given at Section~\ref{sec:illus}. From Proposition~\ref{prop:hmm}, if the family of quantizers  verifies Assumption~\ref{ass:high-rate}, then the conclusions of Theorems~\ref{th:exp_err} and~\ref{th:cv_ordre2} hold true. 

\begin{figure}
\centering
\includegraphics[width=9cm]{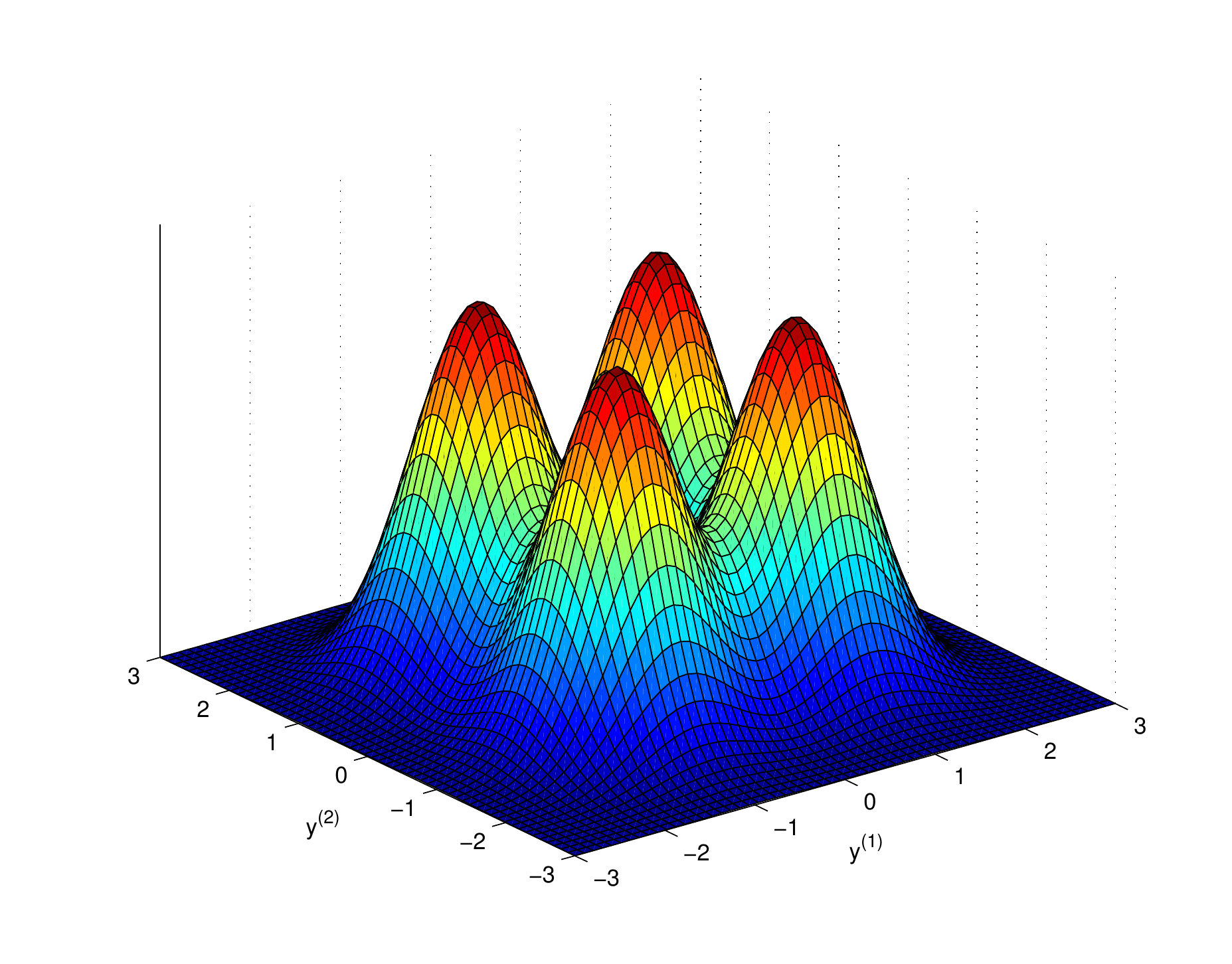}
\caption{QPSK vs. OQPSK --
			Marginal pdf of the observations $p_0(y)=p_1(y)$
			($M=3$, $\sigma=0.6$).}
\label{fig:QPSKvsOQPSK-py}
\end{figure}

Note that the marginal pdf of the measurements $(Y_k)_{k\geq 0}$ (represented in Figure~\ref{fig:QPSKvsOQPSK-py}) writes
\begin{equation}
\label{eq:QPSKvsOQPSK-py}
p_0(y) = p_1(y) = \frac14\,\sum_{x=0}^3 g(x,y) \ .
\end{equation}
Since it does not depend on the hypothesis, Gupta-Hero quantizer~\cite{gupta2003high}, which minimizes the error exponent loss in case of i.i.d. observations, is not defined.

\subsubsection{Examples of Quantizers}

\begin{figure}[!t]
\centering
\begin{tabular}{cc}
\includegraphics[width=7.5cm]{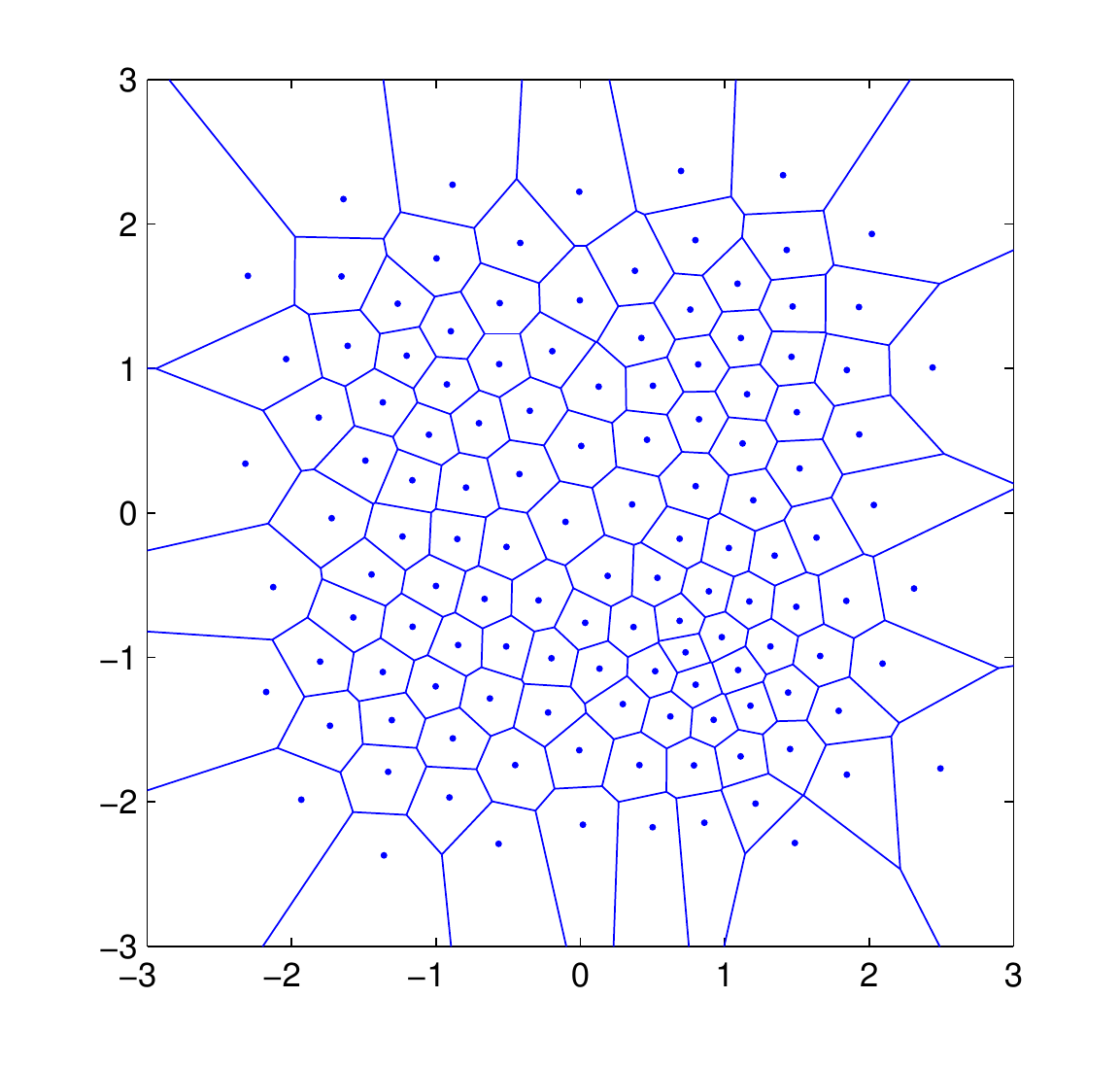}	& \includegraphics[width=7.5cm]{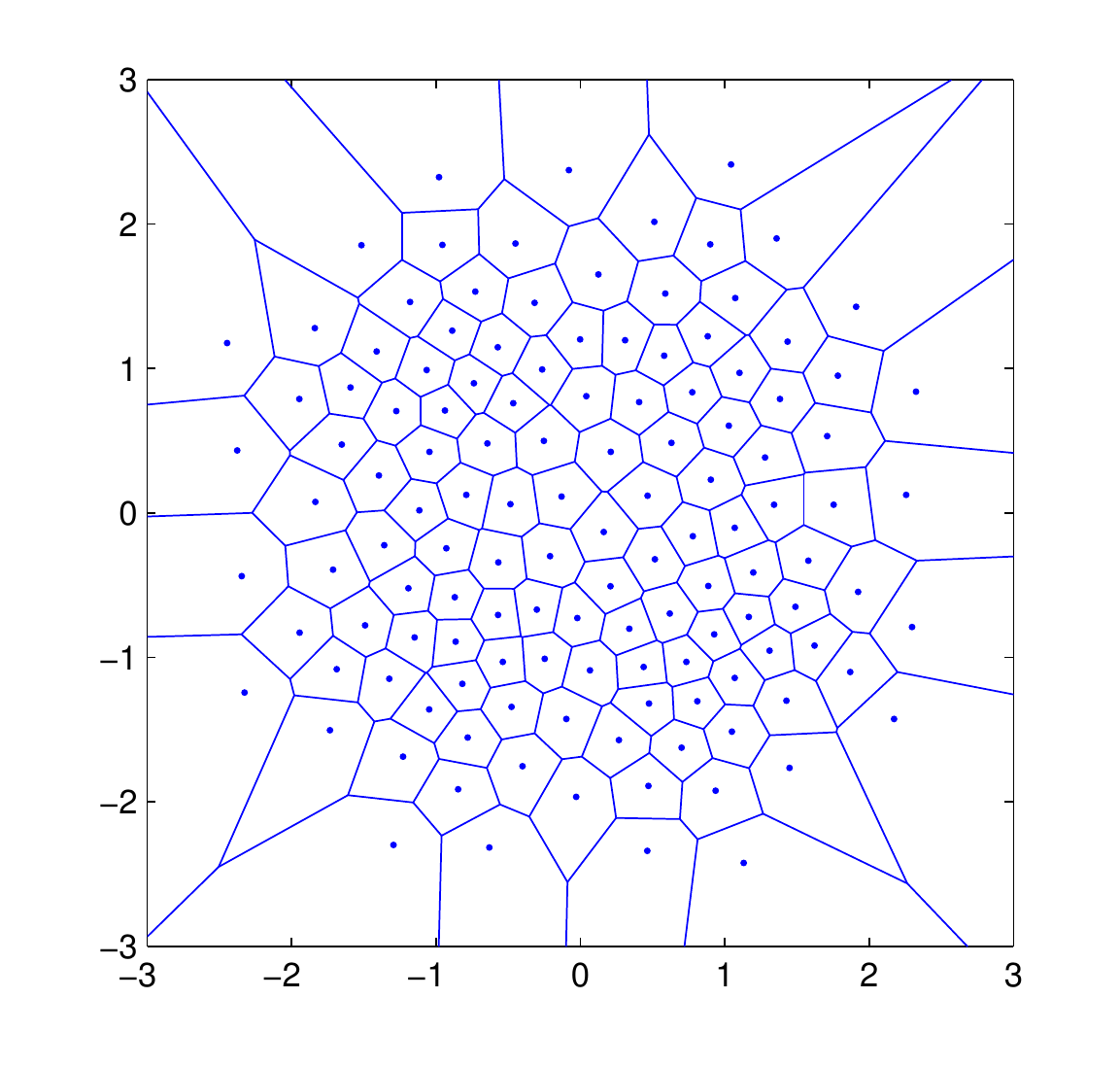}\\
(a)													& (b)
\end{tabular}
\caption{	QPSK vs. OQPSK --
			(a) MSE-optimal $128$-cell quantizer, \
			(b) Proposed $128$-cell quantizer	\
			($M=3$, $\sigma=0.6$, $20\,000$ samples).}
\label{fig:QPSKvsOQPSK-quantizers}
\end{figure}

Figure~\ref{fig:QPSKvsOQPSK-quantizers}(a) represents the MSE-optimal $128$-cell quantizer obtained by the LBG algorithm, and setting $M=3$, $\sigma=0.6$. 
Figure~\ref{fig:QPSKvsOQPSK-quantizers}(b) represents the corresponding proposed quantizer.
Our quantizer is significantly different from the MSE-optimal one. Some low probability points turn out to be significant for the considered detection problem. Details on how we obtained these quantizers are given below.

\paragraph{MSE-optimal quantizer}
The MSE-optimal quantizer of Figure~\ref{fig:QPSKvsOQPSK-quantizers}(a) was obtained by feeding the LBG algorithm with $20\,000$ samples following distribution $\bP_0$ \emph{i.e.}, i.i.d. with pdf $p_0(y)$ (see Figure~\ref{fig:QPSKvsOQPSK-py}).

\paragraph{Proposed quantizer}

As noted in Section~\ref{sec:vector_mini}, the proposed quantizer, whose model point density $\zeta$ is given by Equation~\eqref{eq:LBG_zetaopt}, can be obtained by simply feeding the LBG algorithm with observations corresponding with the following pdf:
\[
q^*(y) = \frac{p_0(y){\bar F}(y)}{\int p_0(s){\bar F}(s)\,ds}\ .
\]
We simulated $20\,000$ samples of this pdf using rejection sampling~\cite[Section~2.2]{liu2001monte}. In practice, we approximated function~$\bar F$ given by Equation~\eqref{eq:Fbarlim} by:
\begin{equation}
\label{eq:approx_Fk}
\bar F_k(y) 
	= \frac1{n_{MC}} \sum_{j=1}^{n_{MC}} \norm{ \nabl{y_0}\log\frac{p_0}{p_1}(Y_{-k:-1}(j),y,Y_{1:k}(j)) }^2 \ ,
\end{equation}
for $k=3$ and $n_{MC}=1\;000$ replications $(Y_m(j))_{m\in\{-k,\dots,-1,1,\dots,k\}, j\in\{1,\dots,n_{MC}\}}$ \emph{i.e.}, $6\;000$ i.i.d. samples with pdf $p_0$. These values were chosen based on empirical observations. 

The gradient in the above equation may be written as follows, after some derivations, and using Equations~\eqref{eq:QPSKvsOQPSK-g},~\eqref{eq:QPSKvsOQPSK-py}:
\begin{eqnarray*}
\nabl{y_0}\log\frac{p_0}{p_1}(y_{-k:k})
	&=& \nabl{y_0}\log p_0(y_0) - \nabl{y_0}\log p_1(y_{-k:k}) \\
	&=& \frac1{\sigma^2}\left\{
			\frac	{\bE_0\left[T(X_0)\,g(X_0,y_0)\right]}
					{\bE_0\left[		g(X_0,y_0)\right]}
		  - \frac	{\bE_1\left[T(X_0)\,\prod_{j=-k}^k g(X_j,y_j)\right]}
					{\bE_1\left[		\prod_{j=-k}^k g(X_j,y_j)\right]}
		\right\} \ .
\end{eqnarray*}
As they are finite sums on $\sX$ or $\sX^{2k+1}$, the above four expectations are exactly computed at the time of the evaluation of $\bar F_k$~\eqref{eq:approx_Fk}.

\subsection{Scenario~\#2: Detection of an AR Structure in Gaussian 2-D Signals}
\label{sec:bbvsAR}

We consider the following model for vector observations with dimension $d=2$:
\[
Y_k = X_k + W_k \ ,
\]
where $W_k \stackrel{i.i.d.}{\sim} \cC\cN(0,\sigma^2)$ represents a zero mean circular Gaussian thermal noise with variance $\sigma^2$,
and where $(X_k)_{k\in\bZ}$ is a Gaussian process which is white under $\HO$
and correlated (AR-1) under $\HU$. 
More precisely,
\[
\begin{array}{l}
  \HO :\  X_k \stackrel{i.i.d.}{\sim} \cC\cN(0,1) \\
  \HU :\  X_k = a X_{k-1} + \sqrt{1-a^2}\,U_k\ ,
\end{array}
\]
where $a\in(0,1)$ is the correlation coefficient and $U_k\stackrel{i.i.d.}{\sim} \cC\cN(0,1)$ is the innovation process.  In particular, $(Y_k)_{k\in\bZ}$ is a white Gaussian process under $\HO$ and is a hidden Markov process under $\HU$, with the particular property that marginal distribution of single observations are identical under both hypotheses.  

We mention that in the above model, densities have infinite support so that the assumptions made in this paper are not satisfied (the observation set $\sY$ coincides with $\bR^2$ and is thus unbounded).
In particular, Theorem~\ref{th:cv_ordre2} does not apply.
Nevertheless, in order to yield some insights on the design of practical quantizers for detection, we can still use the approach described in Section~\ref{sec:vector_mini} and compute the proposed model point density given by Equation~\eqref{eq:LBG_zetaopt}.

\begin{figure}[!t]
\begin{tabular}{cc}
\includegraphics[width=7.5cm]{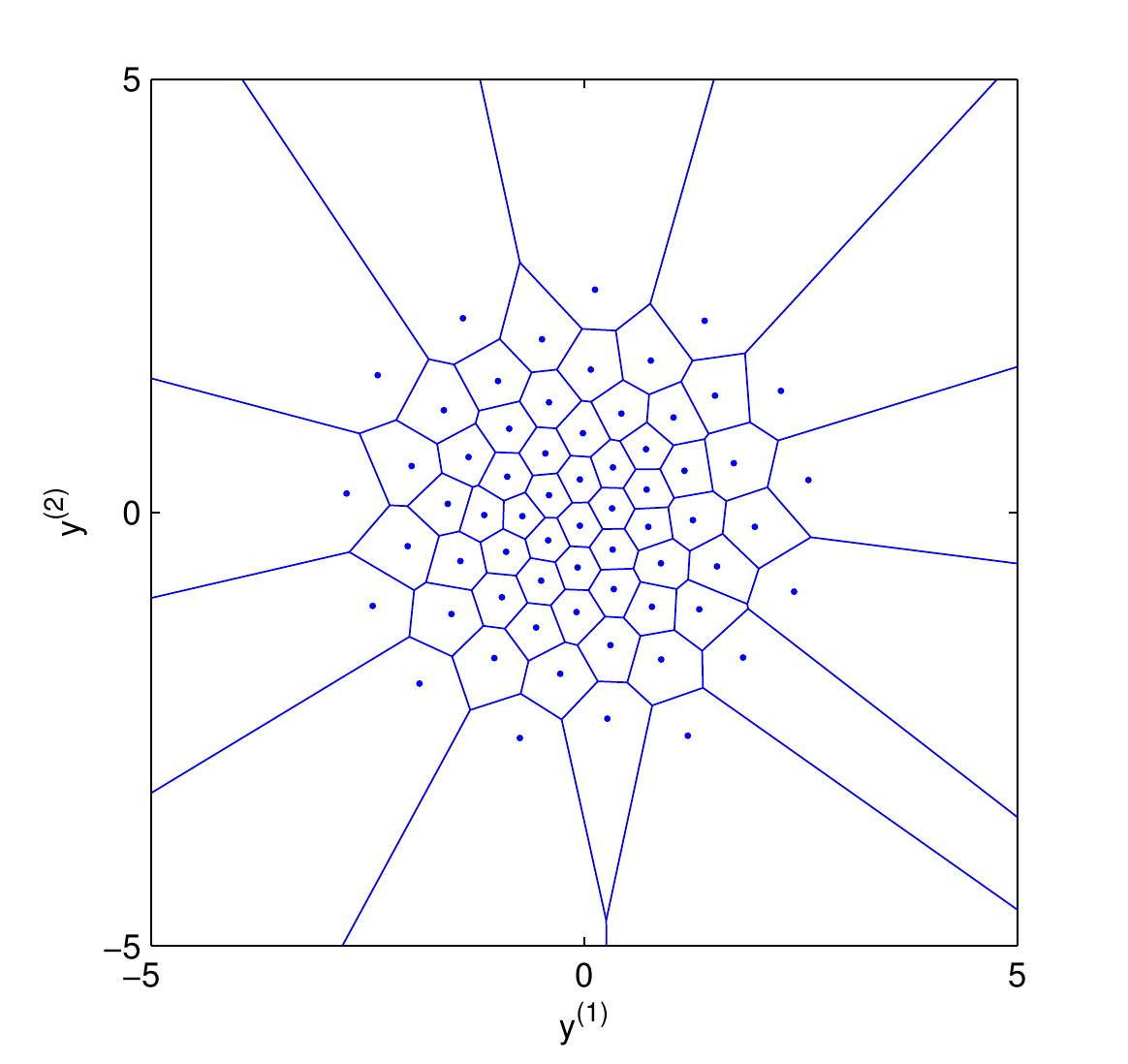}  & \includegraphics[width=7.5cm]{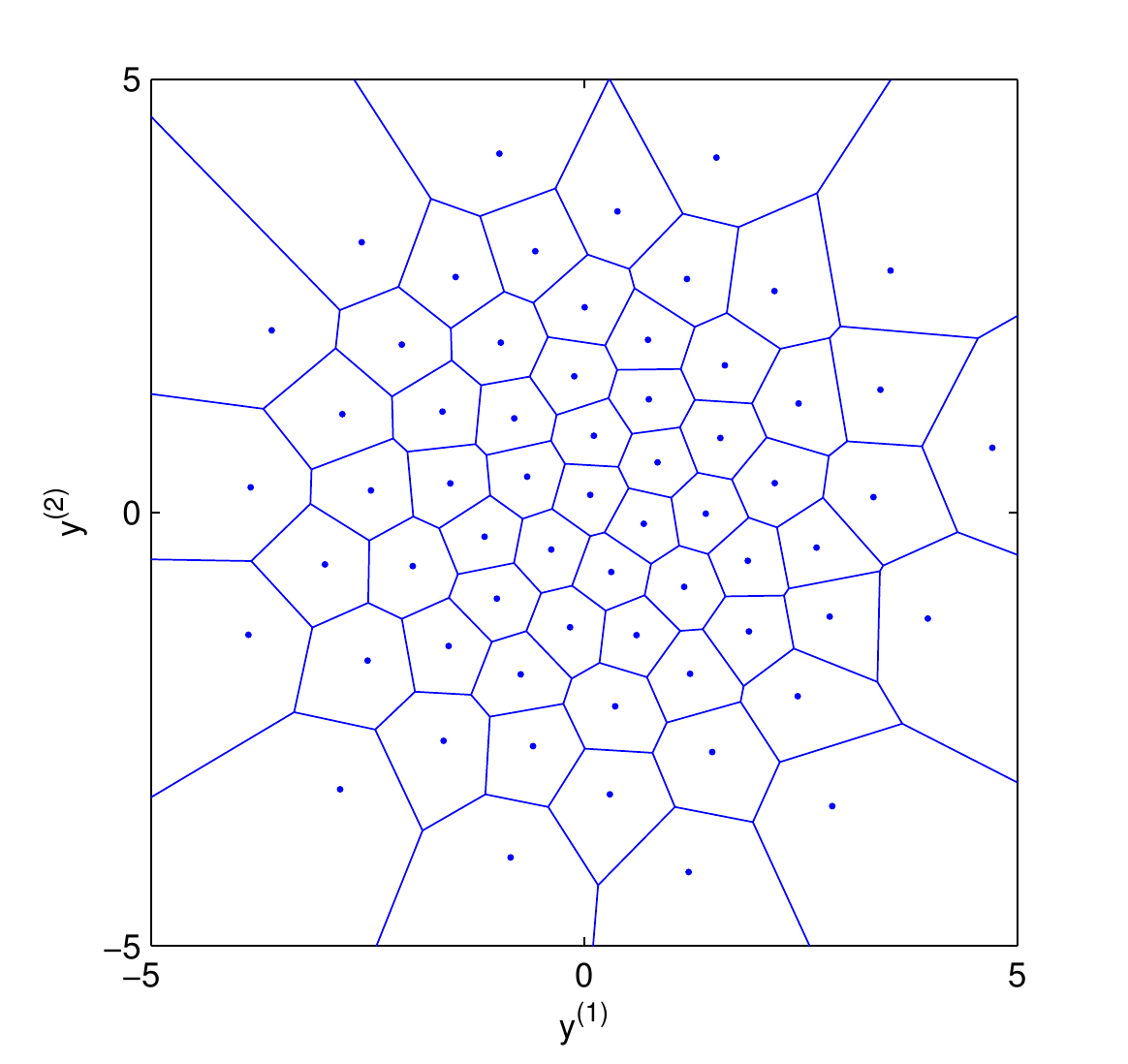}\\
(a)										& (b)
\end{tabular}
\caption{Detection of an AR structure --
			(a) MSE-optimal $64$-cell quantizer, \
			(b) Proposed $64$-cell quantizer	\
			($a=0.8$, $\sigma=1$, $20\,000$ samples).}
\label{fig:bbvsAR}
\end{figure}

Figure~\ref{fig:bbvsAR}(a) represents the MSE-optimal $64$-cell quantizer obtained
by the LBG algorithm (with a $20\,000$-sample training set of data), and setting $\sigma=1$.
Figure~\ref{fig:bbvsAR}(b) represents the corresponding proposed quantizer\footnote{In this case, we approximated function~$\bar F$~\eqref{eq:Fbarlim} for finite $k$ and exactly computed the involved expectation.}, obtained when setting $a=0.8$.
Once again, our quantizer is significantly different from the MSE-optimal one.
As a matter of fact, low probability points seem to be significant for the considered detection problem.

Table~\ref{tab:loss} compares the latter two quantization rules and the uniform one (on the rectangle $[-8;8]^2$) in terms of quantity~$D_e$~\eqref{eq:loss}. As expected, the proposed quantization rule leads to the lowest one. 
We can guess it will also lead to higher detection performance.

\begin{table}[!h]
\caption{Detection of an AR structure --
			Quantity $D_e$ for parameters values $a=0.8$ and $\sigma=1$.}
\label{tab:loss}
\centering
\begin{tabular}{l||c|c|c}
Quantization rule 	& Uniform on $[-8;8]^2$	& MSE-optimal 	& Proposed one	\\
\hline
Quantity $D_e$ 		& 8.211   				& 2.255			& 2.112      
\end{tabular}
\end{table}

\subsection{Scenario~\#3: Detection of a Scalar MA Process in Noise}
\label{sec:0vsMA}

Denote by $Y_k$ the samples collected by a receiver which makes a binary test associated with the following hypotheses:
\[
\begin{array}{l}
  \HO :\  Y_k =  W_k \ ,\\
  \HU :\  Y_k = \displaystyle{\sum_{\ell=0}^L}\,h_\ell\,U_{k-\ell} + W_k \ .
\end{array}
\]
where $W_k \stackrel{i.i.d.}{\sim} \cN(0,\sigma^2)$ represents a thermal noise which is supposed to be real-valued for the sake of illustration.
Here, $U_k$ represents a certain random source which is passed through a propagation channel with deterministic real coefficients $h_0,\dots, h_L$, where $L$ is an integer which represents the channel's memory. 
In the sequel, we set $L=3$.
Assume for instance that $U_k$ is Gaussian distributed $U_k \stackrel{i.i.d.}{\sim} \cN(0,1)$.
We investigate the case where the sensing unit performs a scalar quantization of the received signal before transmission to the decision device.

\begin{figure}[!t]
\centering
\includegraphics[width=9cm]{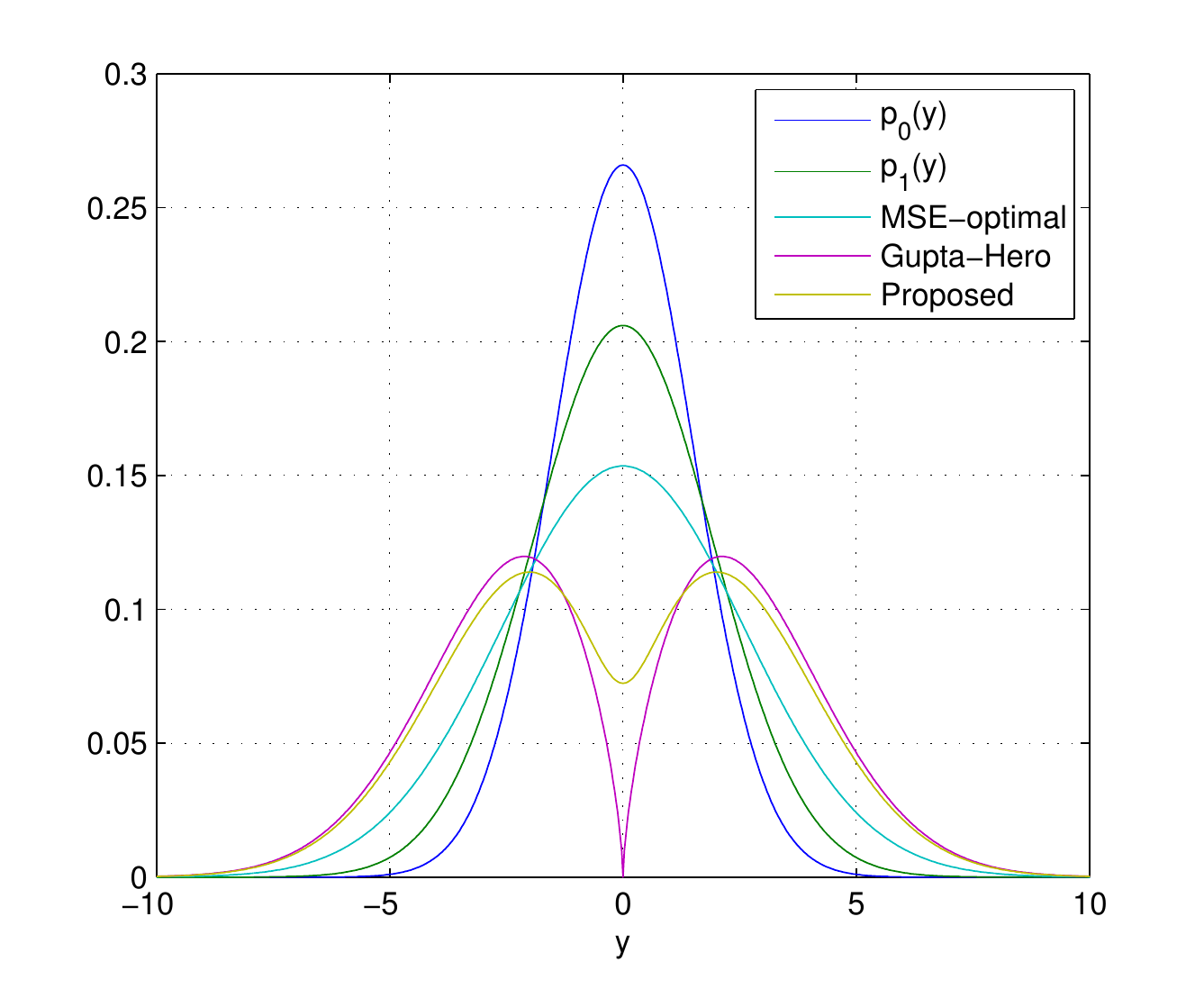}
\caption{Detection of an MA process --
		Probability and model point densities 
		($h=[1.06677,-0.59281,0.09565]$, $\sigma=1.5$).}
\label{fig:0vsMA-densites}
\end{figure}

As in Section~\ref{sec:bbvsAR}, in the above model, densities have infinite support so that the assumptions made in this paper are not satisfied.
Once again, in order to yield some insights on the design of practical quantizers for detection, we can still use the approach described in Section~\ref{sec:vector_mini} and compute the proposed model point density given by Equation~\eqref{eq:LBG_zetaopt}\footnote{In this case, we approximated function~$\bar F$~\eqref{eq:Fbarlim} for finite $k$ and exactly computed the involved expectation.}.

For the same reason, the result of Gupta and Hero~\cite[Equation~(20)]{gupta2003high} does not apply, but we can compute the corresponding quantizer, which model point density is given by~\cite[Equation~(25)]{gupta2003high}, as they did for their Gaussian examples in~\cite[Section~V]{gupta2003high}.

The performance depend on the noise variance $\sigma^2$ and on the particular value of the channel. Thus, we assumed that channel coefficients $h_0,\dots, h_L$ are i.i.d. Gaussian distributed with zero mean and unit variance, and made several simulations.
 
Figure~\ref{fig:0vsMA-densites} represents the probability and model point densities for one channel realization \emph{i.e.}, $h = [1.06677,-0.59281,0.09565]$, and setting $\sigma=1.5$.

\begin{figure}[!h]
\centering
\includegraphics[width=12cm]{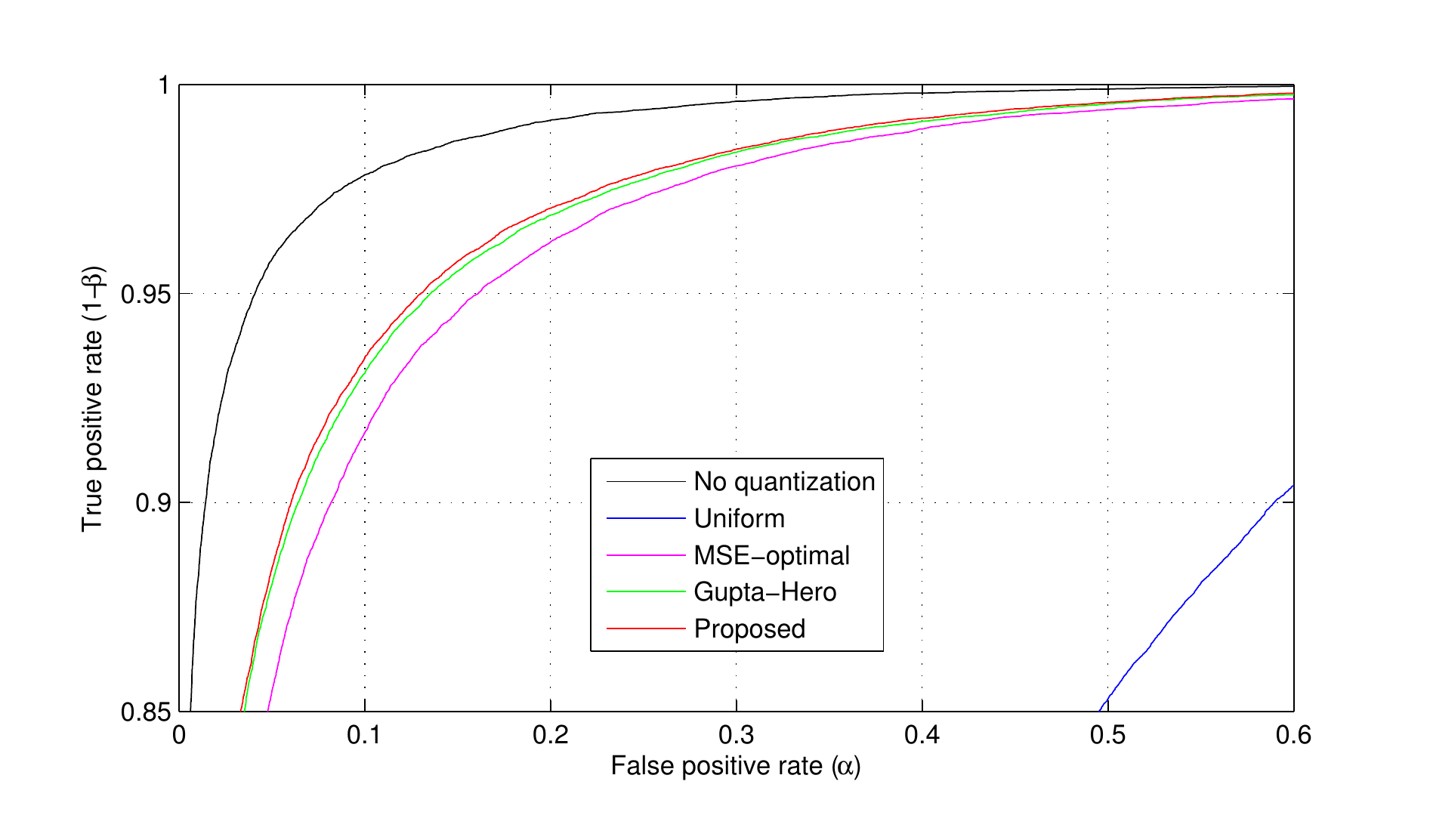}
\caption{Detection of an MA process --
			ROC curves 
			($h=[1.06677,-0.59281,0.09565]$, $\sigma=1.5$, $n=80$, $N=4$, $100\,000$ samples).}
\label{fig:0vsMA-ROC}
\end{figure}

Considering a system with $n=80$ sensors, constructing $4$-cell quantizers for different methods, and computing the corresponding quantized probability distributions under each hypothesis, we can compare the considered quantization rules in terms of detection performance through their respective receiver operating characteristics (ROC curves).
Figure~\ref{fig:0vsMA-ROC} represents such curves for the above channel realization. The uniform quantizer is used on the support $[-10\sigma,10\sigma]$. The whole curve is plotted using $50\,000$ samples of LLR under each hypothesis.

The proposed quantization rule improves the detection performance compared to the MSE-optimal quantizer. In this example, the ROC curve is close to that obtained using Gupta-Hero quantizer. Recall however that in other contexts (\emph{e.g.} in Scenarios~\#1 and \#2), Gupta-Hero quantizer may not even be defined.
We must also qualify this observation: Our theoretical results are valid in the asymptotic regime where $N$ and $n$ tend to infinity, that is,
in the regime where the power of the test tends exponentially to one. 
In practice, the empirical validation of our result would thus require to simulate
rare events. This topic is out of the scope of this paper.

Note that if we interchange $\HO$ and $\HU$, the proposed quantization rule will be different. This is due to the fact that the asymptotic regime we are interested in when dealing with error exponents \emph{i.e.}, $n$ tends to infinity for a fixed type-I error~$\alpha$, restricts attention to one point along the Neyman-Pearson ROC curve.

\section{Conclusion}

We investigated the performance of the Neyman-Pearson detector 
used on quantized versions of a correlated vector-valued stationary process.
It was shown that for a constant false alarm level, the miss probability 
of the test converges exponentially to zero. We determined the error exponent and
we provided a compact and informative expression of the latter in the context
of high-rate quantization. It is proved in particular that when the number $N$
of quantization levels tends to infinity, the error exponent converges at speed $N^{-2/d}$
to the ideal error exponent that one would obtain in the absence of quantization.
In case of scalar quantization, we analytically
characterized the high-rate quantizers minimizing the error exponent loss.
In case of vector quantization, we proposed a method based on the LBG algorithm
in order to construct practical quantizers with attractive performance.

We believe that there are many directions for extending these results and mention a few here. In this paper, observations have absolutely continuous probability distributions w.r.t. the Lebesgue measure. Following Graf and Luschgy~\cite[Section~6]{graf2000foundations} who considered measures with both continuous and singular parts, we could think of an extension of our work to such cases. 

We moreover focused on constant false-alarm rate (CFAR) tests. Following the arguments developed in~\cite{gupta2003high} and using the results of~\cite[Section III]{chen1996general}, it could be interesting to study the whole asymptotic ROC curve and use a global performance criterion like the area under the curve (AUC). However, this would require a nontrivial extension of Sanov's theorem~\cite{dembo1998large} to non-i.i.d. times series.

We furthermore think that the framework developed in this paper could be applied in the context of parameter estimation. The effect of quantization on performance, measured for instance by the Fisher information, could be studied and corresponding optimal vector quantizers could be described.

\appendices

\section{Proof of Lemma~\ref{lem:dev_puN_p1}}
\label{app:dev_puN_p1}

We write the Taylor-Lagrange expansion of function $y_{-m:u} \mapsto p_1(y_{-m:u})$ at point $\xi_{N,j_{-m:u}}$:
\begin{multline}
\label{eq:dev_p1}
p_1(y_{-m:u})	= p_1(\xi_{N,j_{-m:u}}) + \sum_{k=-m}^u \nabl{y_k}p_1(\xi_{N,j_{-m:u}})\T\,(y_k-\xi\Njk) \\
				+ \frac{1}{2} \sum_{k,\ell=-m}^u (y_k-\xi\Njk)\T\,\hess{y_k,y_\ell}p_1(\xi_{N,j_{-m:u}})\,(y_\ell-\xi\Njl)
				+ \epsilon_N(y_{-m:u}) \ ,
\end{multline}
where
\begin{multline*}
\epsilon_N(y_{-m:u}) = \frac{1}{6} \sum_{k,\ell,r=-m}^u \sum_{h,\bar\i,\bar\jmath=1}^d
	(y_k^{(h)}-\xi\Njk^{(h)})(y_\ell^{(\bar\i)}-\xi\Njl^{(\bar\i)})(y_r^{(\bar\jmath)}-\xi\Njr^{(\bar\jmath)}) \\
	×\dtron{p_1}{y_k^{(h)}}{y_\ell^{(\bar\i)}}{y_r^{(\bar\jmath)}}\,(\theta y_{-m:u}+ (1-\theta) \xi_{N,j_{-m:u}}) \ ,
\end{multline*}
for a given $\theta\in[0,1]$ (see~\cite{lang1973calculus}).
Plugging expansion~\eqref{eq:dev_p1} into~\eqref{eq:def_puN} leads to:
\begin{align}
\label{eq:dev_puN_p1}
\frac{\bar p\uN(\xi_{N,j_{-m:u}})}{p_1(\xi_{N,j_{-m:u}})}
	&= 1 + \sum_{k=-m}^u \int_{C\Njk} \frac{\nabl{y_k}p_1(\xi_{N,j_{-m:u}})\T}{p_1(\xi_{N,j_{-m:u}})}\,(y_k-\xi\Njk) \frac{dy_k}{V\Njk}
	\nonumber\\
	&+ \frac12 \sum_{k=-m}^u \int_{C\Njk} (y_k-\xi\Njk)\T\,
			\frac{\hess{y_k}p_1(\xi_{N,j_{-m:u}})}{p_1(\xi_{N,j_{-m:u}})}\,(y_k-\xi\Njk) \frac{dy_k}{V\Njk}
	\nonumber\\
	&+ \frac12 \sum_{k\neq\ell} \int_{C\Njk}\int_{C\Njl} (y_k-\xi\Njk)\T\,
			\frac{\hess{y_k,y_\ell}p_1(\xi_{N,j_{-m:u}})}{p_1(\xi_{N,j_{-m:u}})}\,(y_\ell-\xi\Njl) \frac{dy_k}{V\Njk} \frac{dy_\ell}{V\Njl}
	\nonumber\\
	&+ \epsilon_{N,j_{-m:u}}\ ,
\end{align}
where
\[
\epsilon_{N,j_{-m:u}} = \intm_{C_{N,j_{-m}}×  \cdots ×  C_{N,j_u}}
	\frac{\epsilon_N(y_{-m:u})}{p_1(\xi_{N,j_{-m:u}})}\,\frac{dy_{-m:u}}{\prod_{i=-m}^u V\Nji} \ .
\]

We now determine an estimate for this remainder term. For each
$y_{-m:u} \in C_{N,j_{-m}}×  \cdots ×  C_{N,j_u}$,
\begin{multline}
\label{eq:epsN}
\frac{\epsilon_N(y_{-m:u})}{p_1(\xi_{N,j_{-m:u}})} = \frac{1}{6} \sum_{k,\ell,r=-m}^u \sum_{h,\bar\i,\bar\jmath=1}^d
	(y_k^{(h)}-\xi\Njk^{(h)})(y_\ell^{(\bar\i)}-\xi\Njl^{(\bar\i)})(y_r^{(\bar\jmath)}-\xi\Njr^{(\bar\jmath)}) \\
	× \frac{1}{p_1(\theta y_{-m:u}+ (1-\theta) \xi_{N,j_{-m:u}})}\ 
	\dtron{p_1}{y_k^{(h)}}{y_\ell^{(\bar\i)}}{y_r^{(\bar\jmath)}}\,(\theta y_{-m:u}+ (1-\theta) \xi_{N,j_{-m:u}}) \\
	× \frac{p_1(\theta y_{-m:u}+ (1-\theta) \xi_{N,j_{-m:u}})}{p_1(\xi_{N,j_{-m:u}})} \ .
\end{multline}
First, we find a bound for the last factor. To that end, we expand function
$z_{-m:u} \mapsto \log p_1(z_{-m:u})$ at point $\xi_{N,j_{-m:u}}$:
\[
\log p_1(z_{-m:u}) = \log p_1(\xi_{N,j_{-m:u}}) 
	+ \sum_{k=-m}^u \nabl{y_k} \log p_1(\theta' z_{-m:u}+ (1-\theta')\xi_{N,j_{-m:u}})\T\,(z_k-\xi\Njk) \ ,
\]
for a given $\theta' \in [0,1]$.
From Equation~\eqref{eq:C1}, the following inequality holds:
\begin{align*}
\abs{\log \frac{p_1(z_{-m:u})}{p_1(\xi_{N,j_{-m:u}})}}
	&\leq \sum_{k=-m}^u \norm{\nabl{y_k} \log p_1(\theta' z_{-m:u}+ (1-\theta')\xi_{N,j_{-m:u}})}
					   \norm{z_k-\xi\Njk} \\
	&\leq C_1 \sum_{k=-m}^u \norm{z_k-\xi\Njk} \ .
\end{align*}
Applying the above upper bound at point
$z_{-m:u} = \theta y_{-m:u}+ (1-\theta) \xi_{N,j_{-m:u}}$
and using Assumption~\ref{ass:high-rate}-3), we find 
\[
\abs{\log \frac{p_1(\theta y_{-m:u}+ (1-\theta) \xi_{N,j_{-m:u}})}{p_1(\xi_{N,j_{-m:u}})}}
	\leq C_1 (m+1) \frac{C_d}{N^{1/d}}\ ,
\]
for each $y_{-m:u} \in C_{N,j_{-m}}×  \cdots ×  C_{N,j_u}$.
According to the definition of sequence~$m(N)$ (see Equation~\eqref{eq:mN_zero}), 
the r.h.s. of the above equation vanishes as $N$ tends to infinity. Consequently, the term
$\frac{p_1(\theta y_{-m:u}+ (1-\theta) \xi_{N,j_{-m:u}})}{p_1(\xi_{N,j_{-m:u}})}$
in Equation~\eqref{eq:epsN} is bounded.
This result together with Assumption~\ref{ass:loss}-2) gives the following upper bound:
\[
\abs{\epsilon_{N,j_{-m:u}}} \leq c_T \left(\frac{m+1}{N^{1/d}}\right)^3 \ ,
\]
for some constant $c_T$.

Let us now examine the dominant terms of Equation~\eqref{eq:dev_puN_p1}.
Recall that~$\xi\Nj$ is defined as the centroid of cell~$C\Nj$:
\[
\xi\Nj = \int_{C\Nj} y\,\frac{dy}{V\Nj}\ .
\]
It is straightforward to prove the following two equalities, for any $j\in \{1,\dots,N\}$
and any $d$-by-$d$ matrix $A$:
\begin{align*}
&\int_{C\Nj} (y-\xi\Nj) \frac{dy}{V\Nj} = 0 \ ,\\
&\int_{C\Nj} (y-\xi\Nj)\T A\,(y-\xi\Nj) \frac{dy}{V\Nj} = \tr{A M\Nj} V\Nj^{2/d} \ .
\end{align*}
Plugging above identities in Equation~\eqref{eq:dev_puN_p1}
and recalling that $\zeta\Nj = \frac1{N V\Nj}$
prove Lemma~\ref{lem:dev_puN_p1}.

\section{Proof of Lemma~\ref{lem:dev_UNu}}
\label{app:dev_UNu}

We study each term of the r.h.s. of~\eqref{eq:def_UNu}.  Writing
Taylor-Lagrange expansions of the probability densities and using the
fact that quantization levels are centroids of the cells, we prove the
following three lemmas.  Define function $V_N$ on $\sY$ by
$V_N(y)=V\Nj$ whenever $y\in C\Nj$. 

\begin{lemma}
\label{lem:UNu_terme1}
For each $k\in\{-m,\dots,u\}$, the following equality holds true:
\begin{multline*}
\bE_0\Big[ \nabl{y_k}\log p_1(Z_{N,-m:u})\T\,(Y_k-Z\Nk) \Big] \\
	= \frac1{N^{2/d}}\,\bE_0\Bigg[
			\nabl{y_k}\log p_1(Z_{N,-m:u})\T\,\frac{M_N(Y_k)}{\zeta_N(Y_k)^{2/d}}\,
			\frac{\nabl{y_k}p_0(Y_{-m:k-1},Z\Nk,Y_{k+1:u})}{p_0(Y_{-m:u})} \Bigg]
	+ \bar \epsilon_{N,k} \ ,
\end{multline*}
where $\abs{\bar \epsilon_{N,k}} \leq \frac{c_1'}{N^{3/d}}$ for some constant $c_1'$.
\end{lemma}

\begin{IEEEproof}
We expand the expectation:
\begin{align}
\label{eq:dev_esp1}
\bE_0\Big[ \nabl{y_k}&\log p_1(Z_{N,-m:u})\T(Y_k-Z\Nk) \Big] \nonumber\\
 &= \sum_{j_{-m:u}} \intm_{C_{N,j_{-m}}×  \cdots ×  C_{N,j_u}} \nabl{y_k}\log p_1(\xi_{N,j_{-m:u}})\T\,(y_k-\xi\Njk) p_0(y_{-m:u})\,dy_{-m:u} 	\ .
\end{align}
where $\sum_{j_{-m:u}}$ is a summation over all index vectors $j_{-m:u} \in \{1,\dots, N\}^{u+m+1}$.

For each $j_k\in\{1,\dots, N\}$, we then consider the Taylor-Lagrange expansion of $y_k \mapsto p_0(y_{-m:u})$ at point $\xi\Njk$:
\begin{multline}
\label{eq:dev_p0_1}
p_0(y_{-m:u})
	= p_0(y_{-m:k-1},\xi\Njk,y_{k+1:u}) \\
	+ \nabl{y_k}p_0(y_{-m:k-1},\xi\Njk,y_{k+1:u})\T\,(y_k - \xi\Njk)
	+ \epsilon_{N,k}(y_{-m:u}) \ ,
\end{multline}
where
\[
\epsilon_{N,k}(y_{-m:u})
	= 	(y_k - \xi\Njk)\T\,
		\hess{y_k}p_0(y_{-m:k-1},\theta y_k + (1-\theta)\xi\Njk,y_{k+1:u})\,
		(y_k - \xi\Njk)
\]
for a given $\theta\in[0,1]$. Under Assumption~\ref{ass:loss}-2),
from the counterparts of Equations~\eqref{eq:C1},~\eqref{eq:C2} for density~$p_0$ and
following the argument of Lemma~\ref{lem:dev_puN_p1}
(see Appendix~\ref{app:dev_puN_p1}), we can find a bound for this remainder: For each $y_{-m:u} \in C_{N,j_{-m}}×  \cdots ×  C_{N,j_u}$,
\begin{align}
\label{eq:eps_Nk}
\abs{\epsilon_{N,k}(y_{-m:u})}
	&\leq 	\norm{y_k - \xi\Njk}^2
			\norm{\hess{y_k}p_0(y_{-m:k-1},\theta y_k + (1-\theta)\xi\Njk,y_{k+1:u})}\nonumber\\
	&=  \norm{y_k - \xi\Njk}^2
		\left\lVert\frac{\hess{y_k}p_0(y_{-m:k-1},\theta y_k + (1-\theta)\xi\Njk,y_{k+1:u})}
			{p_0(y_{-m:k-1},\theta y_k + (1-\theta)\xi\Njk,y_{k+1:u})}\right\lVert \nonumber\\
	&\hspace{4cm}
		×	\frac{p_0(y_{-m:k-1},\theta y_k + (1-\theta)\xi\Njk,y_{k+1:u})}{p_0(y_{-m:u})}\,
				p_0(y_{-m:u}) \nonumber\\
	&\leq c\,\norm{y_k - \xi\Njk}^2\,p_0(y_{-m:u}) \ ,
\end{align}
for some constant $c$.

Plugging expansion~\eqref{eq:dev_p0_1} into~\eqref{eq:dev_esp1} leads to 
two dominant terms~$D_{N,1}$ and~$D_{N,2}$ and a remainder~$r_N$:
\[
\bE_0\Big[ \nabl{y_k}\log p_1(Z_{N,-m:u})\T\,(Y_k-Z\Nk) \Big] = D_{N,1}+D_{N,2}+r_N \ .
\]
We successively study each of them. The first dominant term is
\begin{align*} 
D_{N,1}
	&= \sum_{j_{-m:u}} \intm_{C_{N,j_{-m}}×  \cdots ×  C_{N,j_u}} \nabl{y_k}\log p_1(\xi_{N,j_{-m:u}})\T (y_k-\xi\Njk) \\
	&\hspace{7cm}	×  p_0(y_{-m:k-1},\xi\Njk,y_{k+1:u})\,dy_{-m:u} \\
	&= \sum_{j_{-m:u}} \nabl{y_k}\log p_1(\xi_{N,j_{-m:u}})\T \intm_{\{C\Nji\}_{i\neq k}} 
					\left( \int_{C\Njk} (y_k-\xi\Njk) dy_k \right) \\
	&\hspace{7cm}	×  p_0(y_{-m:k-1},\xi\Njk,y_{k+1:u})\,\{dy_i\}_{i\neq k} \\
	&= 0 \ , 
\end{align*}
where $\{dy_i\}_{i\neq k}$ stands for $\prod_{i=-m,i\neq k}^u dy_i$.
The last equality holds true since we have chosen the quantization level~$\xi\Nj$ to be the centroid of cell~$C\Nj$.
\medskip

The second dominant term is
\begin{align}
\label{eq:devp0_2}
D_{N,2} &= \sum_{j_{-m:u}} \intm_{C_{N,j_{-m}}×  \cdots ×  C_{N,j_u}} \nabl{y_k}\log p_1(\xi_{N,j_{-m:u}})\T (y_k-\xi\Njk)(y_k-\xi\Njk)\T \nonumber\\
		& \hspace{7.5cm} ×  \nabl{y_k} p_0(y_{-m:k-1},\xi\Njk,y_{k+1:u})\,dy_{-m:u} \nonumber\\
		&= \sum_{j_{-m:u}} \nabl{y_k}\log p_1(\xi_{N,j_{-m:u}})\T 
			 \intm_{\{C\Nji\}_{i\neq k}} \left( \int_{C\Njk} (y_k-\xi\Njk)(y_k-\xi\Njk)\T dy_k \right) \nonumber\\
		&\hspace{7.5cm}	×  \nabl{y_k}p_0(y_{-m:k-1},\xi\Njk,y_{k+1:u}) \,\{dy_i\}_{i\neq k} \nonumber \\
		&= \sum_{j_{-m:u}} \nabl{y_k}\log p_1(\xi_{N,j_{-m:u}})\T\,M\Njk\,V\Njk^{1+2/d} \nonumber\\
		&\hspace{4cm} ×  \intm_{\{C\Nji\}_{i\neq k}} \nabl{y_k}p_0(y_{-m:k-1},\xi\Njk,y_{k+1:u})\,\{dy_i\}_{i\neq k} \ .
\end{align}
We now write this equality in a simple form.
Obviously, under Assumption~\ref{ass:dens}-2), we can write
\[
\nabl{y_k}p_0(y_{-m:k-1},\xi\Njk,y_{k+1:u})
	= \frac{\nabl{y_k}p_0(y_{-m:k-1},\xi\Njk,y_{k+1:u})}{p_0(y_{-m:u})}\,p_0(y_{-m:u}) \ .
\]
Note that the above expression is independent of $y_k\in C\Nj$, so we can also write
\[
\nabl{y_k}p_0(y_{-m:k-1},\xi\Njk,y_{k+1:u})
	= \int_{C\Nj} \frac{\nabl{y_k}p_0(y_{-m:k-1},\xi\Njk,y_{k+1:u})}{p_0(y_{-m:u})}\, 
			p_0(y_{-m:u})\,\frac{dy_k}{V\Nj} \ .
\]
Equation~\eqref{eq:devp0_2} thus becomes
\begin{align*}
D_{N,2} &=\sum_{j_{-m:u}} \nabl{y_k}\log p_1(\xi_{N,j_{-m:u}})\T\,M\Njk\,V\Njk^{2/d} \\
		&\hspace{3.5cm} × \intm_{\{C\Nji\}} \frac{\nabl{y_k}p_0(y_{-m:k-1},\xi\Njk,y_{k+1:u})}{p_0(y_{-m:u})}\, 
						p_0(y_{-m:u}) \,dy_{-m:u} \\
		&=	\frac1{N^{2/d}}\,\bE_0\left[\nabl{y_k}\log p_1(Z_{N,-m:u})\T \frac{M_N(Y_k)}{\zeta_N(Y_k)^{2/d}}\,
				\frac{\nabl{y_k}p_0(Y_{-m:k-1},Z\Nk,Y_{k+1:u})}{p_0(Y_{-m:u})} \right] \ ,
\end{align*}
where the last line comes from $\zeta\Nj = \frac1{N V\Nj}\,$.

We complete the proof with a bound on the remainder term:
\begin{align*}
\abs{r_N}
	&= \abs{ \sum_{j_{-m:u}} \intm_{C_{N,j_{-m}}×  \cdots ×  C_{N,j_u}} \nabl{y_k}\log p_1(\xi_{N,j_{-m:u}})\T
			(y_k-\xi\Njk)\,\epsilon_{N,k}(y_{-m:u})\ dy_{-m:u} } \\
	&\stackrel{(a)}{\leq}
		C_1\,c \intm \norm{y_k-\xi_N(y_k)}^3\,p_0(y_{-m:u})\,dy_{-m:u} \\
	&\stackrel{(b)}{\leq}
		C_1\,c \left(\frac{C_d}{N^{1/d}}\right)^3 = \frac{c_1'}{N^{3/d}} \ ,
\end{align*}
where inequality~(a) is obtained from Equations~\eqref{eq:C1},~\eqref{eq:eps_Nk}
and~(b) is a consequence of Assumption~\ref{ass:high-rate}-3).

Putting all pieces together proves Lemma~\ref{lem:UNu_terme1}.
\end{IEEEproof}

\begin{lemma}
\label{lem:UNu_terme2_1}
There exists a constant $c_2'$ such that, for each $k\neq\ell\in\{-m,\dots, u\}$,
\[
\bE_0\Big[ (Y_k-Z\Nk)\T\,\hess{y_k,y_\ell}\log p_1(Z_{N,-m:u})\,(Y_\ell-Z\Nl) \Big] \leq \frac{c_2'}{N^{3/d}} \ .
\]
\end{lemma}

\begin{IEEEproof}
For each $k \neq \ell$, we expand the expectation:
\begin{multline}
\label{eq:dev_esp2_1}
\bE_0\left[ (Y_k-Z\Nk)\T\,\hess{y_k,y_\ell}\log p_1(Z_{N,-m:u})\,(Y_\ell-Z\Nl) \right] \\
 = \sum_{j_{-m:u}} \intm_{C_{N,j_{-m}}×  \cdots ×  C_{N,j_u}} (y_k-\xi\Njk)\T\,\hess{y_k,y_\ell}\log p_1(\xi_{N,j_{-m:u}})\,(y_\ell-\xi\Njl)\,
 					p_0(y_{-m:u})\,dy_{-m:u}
\end{multline}
and consider the expansion of $y_k \mapsto p_0(y_{-m:u})$ at point $\xi\Njk$:
\begin{equation}
\label{eq:dev_p0_2}
p_0(y_{-m:u}) = p_0(y_{-m:k-1},\xi\Njk,y_{k+1:u}) + \epsilon_{N,k}'(y_{-m:u}) \ ,
\end{equation}
where, from the counterpart of Equation~\eqref{eq:C1} for density~$p_0$ and
following the argument leading to Equation~\eqref{eq:eps_Nk},
 $\abs{\epsilon_{N,k}'(y_{-m:u})}\leq c'\,\norm{y_k - \xi\Njk}\,p_0(y_{-m:u})$
for some constant $c'$.

Plugging expansion~\eqref{eq:dev_p0_2} into~\eqref{eq:dev_esp2_1} leads to a dominant term and a remainder.
The dominant term is
\begin{align*}
\sum_{j_{-m:u}}& \intm_{C_{N,j_{-m}}×  \cdots ×  C_{N,j_u}} (y_k-\xi\Njk)\T\,\hess{y_k,y_\ell}\log p_1(\xi_{N,j_{-m:u}})\,(y_\ell-\xi\Njl)\\
	&\hspace{8cm}	×  p_0(y_{-m:k-1},\xi\Njk,y_{k+1:u}) \,dy_{-m:u} \\
	&= \sum_{j_{-m:u}} \intm_{\{C\Nji\}_{i\neq k}} \left( \int_{C\Njk} (y_k-\xi\Njk) dy_k \right)
					\hess{y_k,y_\ell}\log p_1(\xi_{N,j_{-m:u}})\,(y_\ell-\xi\Njl) \\
	&\hspace{8cm}	×  p_0(y_{-m:k-1},\xi\Njk,y_{k+1:u}) \,\{dy_i\}_{i\neq k} \\
	&= 0 \ .
\end{align*}

\medskip
Using Equation~\eqref{eq:C2} and Assumption~\ref{ass:high-rate}-3), we find a bound for the remainder term:
\begin{multline}
\label{eq:ineq_C2}
\abs{ \sum_{j_{-m:u}} \intm_{C_{N,j_{-m}}×  \cdots ×  C_{N,j_u}} (y_k-\xi\Njk)\T\,\hess{y_k,y_\ell}\log p_1(\xi_{N,j_{-m:u}})\,(y_\ell-\xi\Njl)\,\epsilon_{N,k}'(y_{-m:u}) \,dy_{-m:u} } \\
	\leq C_2\,c' \left(\frac{C_d}{N^{1/d}}\right)^3 = \frac{c_2'}{N^{3/d}} \ .
\end{multline}

\end{IEEEproof}

\begin{lemma}
\label{lem:UNu_terme2_2}
For each $k\in\{-m,\dots, u\}$,
\begin{multline*}
\bE_0\Big[ (Y_k-Z\Nk)\T\,\hess{y_k}\log p_1(Z_{N,-m:u})\,(Y_k-Z\Nk) \Big] \\
	= \frac1{N^{2/d}}\,\bE_0\left[
		\tr{\hess{y_k}\log p_1(Z_{N,-m:u}) \frac{M_N(Y_k)}{\zeta_N(Y_k)^{2/d}}}
		\frac{p_0(Y_{-m:k-1},Z\Nk,Y_{k+1:u})}{p_0(Y_{-m:u})} \right]
	+ \bar \epsilon_{N,k}' \ ,
\end{multline*}
where $\abs{\bar \epsilon_{N,k}'} \leq \frac{c_2'}{N^{3/d}}$.
\end{lemma}

\begin{IEEEproof}
For each $k$, we expand the expectation:
\begin{multline}
\label{eq:dev_esp2_2}
\bE_0\left[ (Y_k-Z\Nk)\T\,\hess{y_k}\log p_1(Z_{N,-m:u})\,(Y_k-Z\Nk) \right] \\
 = \sum_{j_{-m:u}} \intm_{C_{N,j_{-m}}×  \cdots ×  C_{N,j_u}} (y_k-\xi\Njk)\T\,\hess{y_k}\log p_1(\xi_{N,j_{-m:u}})\,(y_k-\xi\Njk)\,p_0(y_{-m:u})\,dy_{-m:u} \ .
\end{multline}
Plugging expansion~\eqref{eq:dev_p0_2} into~\eqref{eq:dev_esp2_2} leads to a dominant term and a remainder.
The study of the dominant term uses the same arguments as Lemma~\ref{lem:UNu_terme1}.
The final expression comes from the following equality:
\[
\int_{C\Nj} (y-\xi\Nj)\T A\,(y-\xi\Nj)\,dy = \tr{A\,M\Nj}\,V\Nj^{1+2/d} \ ,
\]
for any $d$-by-$d$ matrix $A$, and the definition of the specific
point density $\zeta\Nj = \frac1{N V\Nj}\,$.

Equation~\eqref{eq:ineq_C2} is also valid when $k=\ell$ \emph{i.e.}, for 
the remainder considered here. This proves Lemma~\ref{lem:UNu_terme2_2}.
\end{IEEEproof}

Gathering Equation~\eqref{eq:def_UNu} and
Lemmas~\ref{lem:UNu_terme1}, \ref{lem:UNu_terme2_1}, \ref{lem:UNu_terme2_2} results in
\begin{align*}
U_N(u) = &- \frac1{N^{2/d}} \sum_{k=-m}^u
				\bE_0\left[ \nabl{y_k}\log p_1(Z_{N,-m:u})\T \frac{M_N(Y_k)}{\zeta_N(Y_k)^{2/d}}\,
				\frac{\nabl{y_k}p_0(Y_{-m:k-1},Z\Nk,Y_{k+1:u})}{p_0(Y_{-m:u})} \right] \\
		 &- \frac1{2 N^{2/d}} \sum_{k=-m}^u
				\bE_0\left[\tr{\hess{y_k}\log p_1(Z_{N,-m:u}) \frac{M_N(Y_k)}{\zeta_N(Y_k)^{2/d}}}
				\frac{p_0(Y_{-m:k-1},Z\Nk,Y_{k+1:u})}{p_0(Y_{-m:u})}\right] \\
		 &+ \bar \epsilon_N(u) \ ,
\end{align*}
where $\abs{\bar \epsilon_N(u)} \leq c_U \frac{m^3}{N^{3/d}}$ for some constant $c_U$.

Expanding $\nabl{y_k}p_0$ and $p_0$ once again,
under Assumptions~\ref{ass:high-rate} and \ref{ass:loss}-2),
it is straightforward to write the dominant term in a simple form
\emph{i.e.}, replace each $Z\Nk$ by $Y_k$.
From Equation~\eqref{eq:mN_zero}, the remainder term is a little-o of $N^{-2/d}$.
This proves Lemma~\ref{lem:dev_UNu}.

\section{Proof of Lemma~\ref{lem:series}}
\label{app:series}

Equation~\eqref{eq:nofuture} ensures that the following series converges:
\[
\Sigma_N 	= \bE_0\left[ \cH_{N,0}(Y_{-\infty:0}) \right] 
			+ \sum_{k=-\infty}^{-1} \bE_0\left[\cH\Nk(Y_{-\infty:0}) - \cH\Nk(Y_{-\infty:-1}) \right] \ .
\]
Using Equation~\eqref{eq:K-KN}, the approximation of $N^{2/d}(K-K_N)$
by series $\Sigma_N$ leads to the following remainder:
\begin{equation}
\label{eq:approx_K_KN}
\abs{N^{2/d}(K-K_N) - \Sigma_N} \leq 
		\sum_{k=-m}^{0} \bE_0 \big|\Delta_N^{(k)}\big|
	+ 	\sum_{k=-\infty}^{-m-1} \bE_0 \big|\Upsilon_N^{(k)}\big| 
	+	\check\epsilon_N \ ,
\end{equation}
where
$\Delta_N^{(0)}	=\cH_{N,0}(Y_{-m:0})-\cH_{N,0}(Y_{-\infty:0})$
and
\begin{align*}
  &\Delta_N^{(k)} 	= \cH\Nk(Y_{-m:0}) - \cH\Nk(Y_{-m:-1}) 
					- \cH\Nk(Y_{-\infty:0}) + \cH\Nk(Y_{-\infty:-1}) \quad (\forall\,k\leq -1) \ ,\\
  &\Upsilon_N^{(k)} = \cH\Nk(Y_{-\infty:0}) - \cH\Nk(Y_{-\infty:-1}) \quad (\forall\,k\leq -m-1)\ ,
\end{align*}
and where $\check\epsilon_N\to 0$ as $N\to\infty$.
Using the triangular inequality, we obtain for each $k\leq -1$:
\[
\bE_0 \big|\Delta_N^{(k)}\big| 
	\leq \bE_0 \abs{ \cH\Nk(Y_{-m:0}) - \cH\Nk(Y_{-\infty:0})}
	   + \bE_0 \abs{ \cH\Nk(Y_{-m:-1}) - \cH\Nk(Y_{-\infty:-1})}\ .
\]
Using~\eqref{eq:nopast}, this leads to:
\[
\bE_0\big|\Delta_N^{(k)}\big| \leq 2\,c_h \varphi_{m-\abs{k}}\ .
\]
From the triangular inequality once again,
\[
\bE_0 \big|\Delta_N^{(k)}\big| 
	\leq \bE_0 \abs{ \cH\Nk(Y_{-m:0}) - \cH\Nk(Y_{-m:-1})} 
	   + \bE_0 \abs{ \cH\Nk(Y_{-\infty:0}) -  \cH\Nk(Y_{-\infty:-1})}\ .
\]
Using~\eqref{eq:nofuture}, this leads to:
\[
\bE_0 \big|\Delta_N^{(k)}\big| \leq 2\,c_h \psi_{\abs{k}}\ .
\]
After some algebra, there exists a constant $c_\Delta$ such that:
\begin{align*}
\sum_{k=-m}^{-1} \bE_0 \big|\Delta_N^{(k)}\big|
	&\leq c_\Delta  \sum_{k=-m}^{-1}  \varphi_{m-\abs{k}} \land  \psi_{\abs{k}} \\
	&\leq c_\Delta  \left(\sum_{k=-\floor{m/2}}^{-1}  \varphi_{m-\abs{k}} + \sum_{k=-m}^{-\floor{m/2}}  \psi_{\abs{k}} \right) \\
	&\leq c_\Delta  \left(\sum_{k=\ceil{m/2}}^{\infty}  \varphi_k + \sum_{k=\floor{m/2}}^{\infty}  \psi_k \right) \\
	&\leq c_\Delta\,\cT_\Delta^{(m)}\ ,
\end{align*}
Where $(\cT_\Delta^{(m)})_{m\geq 0}$ is a sequence of positive numbers
such that $\cT_\Delta^{(m)}\to 0$ as $m\to\infty$. 
The last line of the above equation holds true under Assumption~\ref{ass:loss}-4)
since $\sum\varphi_k$ and $\sum\psi_k$ are convergent series. 
Similarly, $\bE_0 \big|\Delta_N^{(0)}\big|\leq \,c_h \varphi_m$.

The last series in~\eqref{eq:approx_K_KN} can be bounded using~\eqref{eq:nofuture}:
\[
\sum_{k=-\infty}^{-m-1} \bE_0 \big|\Upsilon_N^{(k)}\big|
	\leq c_h \sum_{k=-\infty}^{-m-1} \psi_{\abs{k}}  = c_\Upsilon\,\cT_\Upsilon^{(m)} \ ,
\]
for some constant $c_\Upsilon$ and a given sequence $(\cT_\Upsilon^{(m)})_{m\geq 0}$
such that $\cT_\Upsilon^{(m)}\to 0$ as $m\to\infty$.

Putting all pieces together, Equation~\eqref{eq:approx_K_KN} leads to:
\[
\abs{N^{2/d}(K-K_N) - \Sigma_N}
	\leq c_h \varphi_m
	+ c_\Delta\,\cT_\Delta^{(m)}
	+ c_\Upsilon\,\cT_\Upsilon^{(m)} 
	+ \check\epsilon_N \ .
\]
The r.h.s. of the above inequality tends to zero as $m,N\to\infty$. This proves Lemma~\ref{lem:series}.

\section*{Acknowledgment}

The authors would like to thank Prof. Eric Moulines for helpful comments and for bringing useful references to their attention. They are also grateful to Dr. Walid Hachem and Dr. Pablo Piantanida for fruitful discussions.

\bibliographystyle{IEEEtran}
\bibliography{quantif}

\begin{thebibliography}{10}
\providecommand{\url}[1]{#1}
\csname url@samestyle\endcsname
\providecommand{\newblock}{\relax}
\providecommand{\bibinfo}[2]{#2}
\providecommand{\BIBentrySTDinterwordspacing}{\spaceskip=0pt\relax}
\providecommand{\BIBentryALTinterwordstretchfactor}{4}
\providecommand{\BIBentryALTinterwordspacing}{\spaceskip=\fontdimen2\font plus
\BIBentryALTinterwordstretchfactor\fontdimen3\font minus
  \fontdimen4\font\relax}
\providecommand{\BIBforeignlanguage}[2]{{%
\expandafter\ifx\csname l@#1\endcsname\relax
\typeout{** WARNING: IEEEtran.bst: No hyphenation pattern has been}%
\typeout{** loaded for the language `#1'. Using the pattern for}%
\typeout{** the default language instead.}%
\else
\language=\csname l@#1\endcsname
\fi
#2}}
\providecommand{\BIBdecl}{\relax}
\BIBdecl

\bibitem{akyildiz2002wireless}
I.~Akyildiz, W.~Su, Y.~Sankarasubramaniam, and E.~Cayirci, ``Wireless sensor
  networks: a survey,'' \emph{Computer Networks}, vol.~38, no.~4, pp. 393--422,
  2002.

\bibitem{chen2006channel}
B.~Chen, L.~Tong, and P.~Varshney, ``Channel-aware distributed detection in
  wireless sensor networks,'' \emph{IEEE Signal Process. Mag.}, vol. 1053, no.
  5888/06, pp. 16--26, 2006.

\bibitem{gray1998quantization}
R.~Gray and D.~Neuhoff, ``Quantization,'' \emph{IEEE Trans. Inf. Theory},
  vol.~44, no.~6, pp. 2325--2383, 1998.

\bibitem{gersho1992vector}
A.~Gersho and R.~Gray, \emph{Vector quantization and signal compression}.\hskip
  1em plus 0.5em minus 0.4em\relax Kluwer, 1992.

\bibitem{bennett1948spectra}
W.~Bennett, ``Spectra of quantized signals,'' \emph{Bell System Technical
  Journal}, vol.~27, pp. 446--472, 1948.

\bibitem{na1995bennett}
S.~Na and D.~Neuhoff, ``{Bennett's} integral for vector quantizers,''
  \emph{IEEE Trans. Inf. Theory}, vol.~41, no.~4, pp. 886--900, 1995.

\bibitem{han1998statistical}
T.~Han and S.~Amari, ``Statistical inference under multiterminal data
  compression,'' \emph{IEEE Trans. Inf. Theory}, vol.~44, no.~6, pp.
  2300--2324, 1998.

\bibitem{misra2008distributed}
V.~Misra, V.~Goyal, and L.~Varshney, ``Distributed functional scalar
  quantization: High-resolution analysis and extensions,'' \emph{Arxiv}, vol.
  cs.IT, p. arXiv:0811.3617, 2008.

\bibitem{xiao2006distributed}
J.-J. Xiao, A.~Ribeiro, Z.-Q. Luo, and G.~Giannakis, ``Distributed
  compression-estimation using wireless sensor networks,'' \emph{IEEE Signal
  Process. Mag.}, vol.~23, no.~4, pp. 27--41, 2006.

\bibitem{perlmutter1996bayes}
K.~Perlmutter, S.~Perlmutter, R.~Gray, R.~Olshen, and K.~Oehler, ``{Bayes} risk
  weighted vector quantization with posterior estimation for image compression
  and classification,'' \emph{IEEE Trans. Image Process.}, vol.~5, no.~2, pp.
  347--360, 1996.

\bibitem{kassam1977optimum}
S.~Kassam, ``Optimum quantization for signal detection,'' \emph{IEEE Trans.
  Commun.}, vol.~25, no.~5, pp. 479--484, 1977.

\bibitem{poor1977applications}
H.~Poor and J.~Thomas, ``Applications of {Ali--Silvey} distance measures in the
  design of generalized quantizers for binary decision systems,'' \emph{IEEE
  Trans. Commun.}, vol.~25, no.~9, pp. 893--900, 1977.

\bibitem{poor1988fine}
H.~Poor, ``Fine quantization in signal detection and estimation,'' \emph{IEEE
  Trans. Inf. Theory}, vol.~34, no.~5, pp. 960--972, 1988.

\bibitem{picinbono1988optimum}
B.~Picinbono and P.~Duvaut, ``Optimum quantization for detection,'' \emph{IEEE
  Trans. Commun.}, vol.~36, no.~11, pp. 1254--1258, 1988.

\bibitem{tsitsiklis1993extremal}
J.~Tsitsiklis, ``Extremal properties of likelihood-ratio quantizers,''
  \emph{IEEE Trans. Commun.}, vol.~41, no.~4, pp. 550--558, 1993.

\bibitem{tenney1981detection}
R.~Tenney and N.~Sandell, ``Detection with distributed sensors,'' \emph{IEEE
  Transactions on Aerospace and Electronic Systems}, vol.~17, no.~4, pp.
  501--510, 1981.

\bibitem{tsitsiklis1988decentralized}
J.~Tsitsiklis, ``Decentralized detection by a large number of sensors,''
  \emph{Mathematics of Control, Signals, and Systems}, vol.~1, no.~2, pp.
  167--182, 1988.

\bibitem{gupta2003high}
R.~Gupta and A.~Hero, ``High-rate vector quantization for detection,''
  \emph{IEEE Trans. Inf. Theory}, vol.~49, no.~8, pp. 1951--1969, 2003.

\bibitem{lehmann2005testing}
E.~Lehmann and J.~Romano, \emph{Testing Statistical Hypotheses (3rd Ed)}.\hskip
  1em plus 0.5em minus 0.4em\relax Springer Texts in Statistics, 2005.

\bibitem{cover2006elements}
T.~Cover and J.~Thomas, \emph{Elements of information theory (2nd Ed)}.\hskip
  1em plus 0.5em minus 0.4em\relax Wiley-Interscience, 2006.

\bibitem{chamberland2006dense}
J.-F. Chamberland and V.~Veeravalli, ``How dense should a sensor network be for
  detection with correlated observations?'' \emph{IEEE Trans. Inf. Theory},
  vol.~52, no.~11, pp. 5099--5106, 2006.

\bibitem{willett2000good}
P.~Willett, P.~Swaszek, and R.~Blum, ``The good, bad, and ugly: Distributed
  detection of a known signal in dependent {Gaussian} noise,'' \emph{IEEE
  Trans. Signal Process.}, vol.~48, no.~12, pp. 3266--3279, 2000.

\bibitem{sung2006neyman}
Y.~Sung, L.~Tong, and H.~Poor, ``{Neyman--Pearson} detection of {Gauss--Markov}
  signals in noise : closed-form error exponent and properties,'' \emph{IEEE
  Trans. Inf. Theory}, vol.~52, no.~4, pp. 1354--1365, 2006.

\bibitem{hachem2009error}
W.~Hachem, E.~Moulines, and F.~Roueff, ``Error exponents for {Neyman--Pearson}
  detection of a continuous-time {Gaussian Markov} process from noisy irregular
  samples,'' \emph{arXiv cs.IT}, 2009, submitted to IEEE Trans. on Inf. Theory.

\bibitem{chen1996general}
P.-N. Chen, ``General formulas for the {Neyman--Pearson} type-{II} error
  exponent subject to fixed and exponential type-{I} error bounds,'' \emph{IEEE
  Trans. Inf. Theory}, vol.~42, no.~1, pp. 316--323, 1996.

\bibitem{bradley2005basic}
R.~Bradley, ``Basic properties of strong mixing conditions. a survey and some
  open questions,'' \emph{Probability Surveys}, vol.~2, pp. 107--144, 2005.

\bibitem{moy1961generalizations}
S.~Moy, ``Generalizations of {Shannon--McMillan} theorem,'' \emph{Pacific J.
  Math.}, vol.~11, no.~2, pp. 705--714, 1961.

\bibitem{doukhan1994mixing}
P.~Doukhan, \emph{Mixing: properties and examples}.\hskip 1em plus 0.5em minus
  0.4em\relax Springer, 1994.

\bibitem{bosq1998nonparametric}
D.~Bosq, \emph{Nonparametric statistics for stochastic processes: estimation
  and prediction}.\hskip 1em plus 0.5em minus 0.4em\relax Springer Verlag,
  1998.

\bibitem{villardISIT}
J.~Villard and P.~Bianchi, ``High-rate vector quantization for the
  {Neyman-Pearson} detection of some stationary mixing processes,'' in
  \emph{ISIT}, Austin, Texas, USA, 2010.

\bibitem{panter1951quantization}
P.~Panter and W.~Dite, ``Quantization distortion in pulse-count modulation with
  nonuniform spacing of levels,'' \emph{Proceedings of the IRE}, vol.~39,
  no.~1, pp. 44 -- 48, 1951.

\bibitem{neuhoff1996asymptotic}
D.~Neuhoff, ``On the asymptotic distribution of the errors in vector
  quantization,'' \emph{IEEE Trans. Inf. Theory}, vol.~42, no.~2, pp. 461--468,
  1996.

\bibitem{gersho1979asymptotically}
A.~Gersho, ``Asymptotically optimal block quantization,'' \emph{IEEE Trans.
  Inf. Theory}, vol.~25, no.~4, pp. 373--380, 1979.

\bibitem{zamir1996lattice}
R.~Zamir and M.~Feder, ``On lattice quantization noise,'' \emph{IEEE
  Transactions on Information Theory}, vol.~42, no.~4, pp. 1152 --1159, 1996.

\bibitem{linde1980algorithm}
Y.~Linde, A.~Buzo, and R.~Gray, ``An algorithm for vector quantizer design,''
  \emph{IEEE Trans. Commun.}, vol.~28, no.~1, pp. 84--95, 1980.

\bibitem{conway1999sphere}
J.~Conway and N.~Sloane, \emph{Sphere packings, lattices, and groups (3rd
  Ed)}.\hskip 1em plus 0.5em minus 0.4em\relax Springer-Verlag, 1999.

\bibitem{gupta2001quantization}
R.~Gupta, ``Quantization strategies for low-power communications,'' Ph.D.
  dissertation, The University of Michigan, 2001.

\bibitem{lasserre1995trace}
J.~Lasserre, ``A trace inequality for matrix product,'' \emph{IEEE Trans.
  Autom. Control}, vol.~40, no.~8, pp. 1500 --1501, 1995.

\bibitem{marshall1979inequalities}
A.~Marshall and I.~Olkin, \emph{Inequalities: theory of majorization and its
  applications}.\hskip 1em plus 0.5em minus 0.4em\relax Academic Press New
  York, 1979.

\bibitem{billingsley1995probability}
P.~Billingsley, \emph{Probability and Measure (3rd Ed)}.\hskip 1em plus 0.5em
  minus 0.4em\relax John Wiley \& Sons, 1995.

\bibitem{douc2004asymptotic}
R.~Douc, E.~Moulines, and T.~Ryden, ``Asymptotic properties of the maximum
  likelihood estimator in autoregressive models with {Markov} regime,''
  \emph{The Annals of Statistics}, vol.~32, no.~5, pp. 2254--2304, 2004.

\bibitem{cappe2007inference}
O.~Cappé, E.~Moulines, and T.~Ryden, \emph{Inference in Hidden {Markov}
  Models}.\hskip 1em plus 0.5em minus 0.4em\relax Springer series in
  statistics, 2007.

\bibitem{proakis2007digital}
J.~Proakis and M.~Salehi, \emph{Digital communications (5th Ed)}.\hskip 1em
  plus 0.5em minus 0.4em\relax McGraw-Hill, 2007.

\bibitem{johnson1994continuous}
N.~Johnson, S.~Kotz, and N.~Balakrishnan, \emph{Continuous univariate
  distributions, vol. 1 (2nd Ed)}.\hskip 1em plus 0.5em minus 0.4em\relax
  Wiley-Interscience, 1994.

\bibitem{liu2001monte}
J.~Liu, \emph{{Monte Carlo} strategies in scientific computing}.\hskip 1em plus
  0.5em minus 0.4em\relax Springer Verlag, 2001.

\bibitem{graf2000foundations}
S.~Graf and H.~Luschgy, \emph{Foundations of quantization for probability
  distributions}.\hskip 1em plus 0.5em minus 0.4em\relax Springer, 2000.

\bibitem{dembo1998large}
A.~Dembo and O.~Zeitouni, \emph{Large deviations techniques and applications
  (2nd Ed)}.\hskip 1em plus 0.5em minus 0.4em\relax Springer Verlag, 1998.

\bibitem{lang1973calculus}
S.~Lang, \emph{Calculus of several variables}.\hskip 1em plus 0.5em minus
  0.4em\relax Addison-Wesley, 1973.

\end{thebibliography}

\begin{IEEEbiographynophoto}
{Joffrey Villard} (S'09) was born in Saint-{\'E}tienne, France, in 1985.
He received the Dipl.Ing. degree in digital communication and electronics, and the M.Sc. degree in wireless communication systems, both from Sup{\'e}lec, Gif-sur-Yvette, France, in 2008. 

He is currently working towards the Ph.D. degree at the Department of Telecommunications of Sup{\'e}lec.
His research interests include information theory, source coding, statistical inference, and signal processing for wireless sensor networks.
\end{IEEEbiographynophoto}

\begin{IEEEbiographynophoto}
{Pascal Bianchi} (M'06) was born in 1977 in Nancy, France.
He received the M.Sc. degree of Sup{\'e}lec-Paris XI in 2000 and the Ph.D. degree of the University of Marne-la-Vall{\'e}e in 2003. 

From 2003 to 2009, he was an Associate Professor at the Telecommunication Department of Supélec.
In 2009, he joined the Statistics and Applications group at LTCI-Telecom ParisTech.
His current research interests are in the area of statistical signal processing for sensor networks.
They include decentralized detection, quantization, stochastic optimization, and applications of random matrix theory.
\end{IEEEbiographynophoto}

\vfill
\end{document}